\documentclass[preprint,eqsecnum,aps]{revtex4}

\usepackage{graphicx}
\usepackage{dcolumn}
\usepackage{amsmath}

\begin{document}

\title{Optical properties of unconventional superconductors}

\author{Takashi Yanagisawa$^{a,b}$ and Hajime Shibata$^a$} 

\affiliation{$^a$Nanoelectronics Research Institute (NeRI), National Institute of 
Advanced Industrial Science and Technology (AIST), Central 2 1-1-1 Umezono 
Tsukuba 305-8568, Japan\\
$^b$Graduate School of Science, Osaka University, Toyonaka 560-0043, Japan}

\date{\today}

\begin{abstract}
The optical conductivity measurements give a powerful tool to
investigate the nature of the superconducting gap for conventional and 
unconventional superconductors.  In this article, first, general analyses of the 
optical 
conductivity are given stemmed from the Mattis-Bardeen formula for conventional
BCS superconductors to unconventional anisotropic superconductors.  Second, we 
discuss the reflectance-transmittance (R-T) method which has been 
proposed to measure far-infrared spectroscopy.  The R-T method provides us 
precise measurements of the frequency-dependent conductivity.  Third, the 
optical conductivity spectra of the electron-doped cuprate superconductor 
Nd$_{2-x}$Ce$_x$CuO$_4$ are investigated based on the anisotropic pairing model.
It is shown that the behavior of optical conductivity is consistent with an 
anisotropic gap and is well explained by
the formula for d-wave pairing in the far-infrared region.  
The optical properties of the multiband superconductor MgB$_2$, in which the 
existence of superconductivity with relatively high-$T_c$ (39K) was recently 
announced, is also examined to determine the symmetry of superconducting gaps.
\end{abstract}

\maketitle

 
\section{Introduction}
The measurements of optical properties provide us important insights concerning
the nature of charge carries, pseudogaps and superconducting gaps, as well as
the electronic band structure of a material.\cite{tim89}
The optical spectroscopy gives a view into the electronic structure, low-lying
excitations, phonon structure, etc. 
The optical conductivity or the dielectric function indicates a response of a 
system of electrons to an applied field.
For the ordinary superconductors the evidence for an energy gap has been obtained
by infrared spectroscopy.  Far above the superconducting energy gap, a bulk
superconductor behaves like a normal metal in the optical response.
The Mattis-Bardeen formula derived in the BCS theory consistently describes the
infrared behaviors in the classical conventional superconductors.
After the discovery of high-temperature superconductivity, a large amount of
works has been made to find the superconducting gap and any spectral features
responsible to the superconducting pairing, using an infrared spectroscopy
technique.

Optical properties are discussed in the linear response theory where the induced
currents are proportional to the external applied field.  General formulas have
been derived for the optical response.
In this paper in Section II we discuss the linear response theory for the conductivity;
we derive the Mattis-Bardeen formula for conventional superconductors and the
formula for London superconductors.  
The conductivity sum rule is briefly discussed here.

In Section III we briefly present a new method to characterize far-infrared
optical properties which we call the reflectance-transmittance method (R-T method).
In this method, both the reflectance spectra $R(\omega)$ and the transmittance
spectra $T(\omega)$ are measured, and then they are substituted into a set of coupled
equations which describe exactly the transmittance and reflectance of thin films.
The coupled equations are solved numerically by the Newton method to obtain the
complex refractive indices $n$ and $k$ of thin films as functions of the frequency
$\omega$, which determined the optical conductivity $\sigma(\omega)$.
Since this method does not need a Kramers-Kronig transformation, we are free from
difficulties stemmed from uncertainties in the small $\omega$ region in the 
conventional method.

In the subsequent Sections we discuss two materials: the electron-doped oxide
superconductor Nd$_{2-x}$Ce$_x$CuO$_4$ and the magnesium diboride MgB$_2$ exhibiting
$T_c=39$K.  The cuprate high-$T_c$ superconductors are regarded as a typical London
superconductor satisfying $\lambda \gg \xi$ for the penetration depth and the
coherence length.
Our date obtained from the R-T method for Nd$_{2-x}$Ce$_x$CuO$_4$ clearly indicates 
the $d$-wave symmetry with nodes for the superconducting gap.

MgB$_2$ is recently discovered superconductor with a relatively high $T_c$ in spite
of its simple crystal structure.
The symmetry of Cooper pairs is an issue which should be clarified to investigate
the mechanism of high $T_c$.  The optical properties provide us information on
superconducting gaps from which we conclude that this material is described by
two order parameters attached to $\sigma$- and $\pi$-bands.  Besides, the two order
parameters have different anisotropy to explain the experimental results
consistently.

\section{Theory of Optical Conductivity}
\subsection{Linear Response Theory}
In this section we discuss the optical properties in the linear response theory
in the normal metal and superconductors.  The famous Mattis-Bardeen formula is
derived and its modifications to unconventional superconductors are discussed.
In the Kubo theory the external applied field
\begin{equation}
H'(t)= -\sum_{\mu}a_{\mu}X_{\mu}(t)
=-\sum_{\mu}a_{\mu}X_{\mu}{\rm e}^{-i\omega t},
\end{equation}
is considered as a perturbation to the non-interacting system described by the
Hamiltonian $H_0$.  The total Hamiltonian is given by $H=H_0+H'$. 
From the equation $i\hbar\partial\rho/\partial t=[H,\rho]$, a linear variation 
$\rho'(t)$ to the density operator
$\rho_0={\rm e}^{-\beta H_0}/Z_0$ is written as
\begin{equation}
i\hbar\frac{\partial\rho'}{\partial t}=[H_0,\rho']+[H',\rho_0].
\end{equation}
Then $\rho'$ is given as
\begin{equation}
\rho'(t)= 
-\frac{i}{\hbar}\int^t_{-\infty}dt' {\rm e}^{-iH_0(t-t')/\hbar}
[H'(t'),\rho_0]{\rm e}^{iH(t-t')/\hbar}.
\end{equation}
For the electrical conductivity, the external fields are given by
\begin{equation}
a_{\mu}= e\sum_i x_{i\mu},~~X_{\mu}=E_{\mu}~~(\mu=x,y,z).
\end{equation}
The current is given by
\begin{equation}
J_{\mu}=\dot{a}_{\mu}= e\sum_i(\dot{x}_i)_{\mu}= e\sum_i(v_i)_{\mu}.
\end{equation}
The expectation value of the current is
\begin{eqnarray}
\langle J_{\mu}(t)\rangle&=& {\rm Tr}\rho' J_{\mu}\nonumber\\
&=& -\frac{i}{\hbar}\int^t_{-\infty}dt'{\rm Tr}{\rm e}^{-iH_0(t-t')/\hbar}
[H'(t'),\rho_0]{\rm e}^{iH_0(t-t')/\hbar}J_{\mu}\nonumber\\
&=& \frac{i}{\hbar}\int^{\infty}_0 d\tau {\rm e}^{i\omega\tau}{\rm Tr}
{\rm e}^{-iH_0\tau/\hbar}[a_{\nu},\rho_0]{\rm e}^{iH_0\tau/\hbar}E_{\nu}
{\rm e}^{-i\omega t}J_{\mu}.
\end{eqnarray}
We assume the time dependence of $\langle J_{\mu}(t)\rangle$ as
$\langle J_{\mu}(t)\rangle=J_{\mu}(\omega){\rm e}^{-i\omega t}$, then we obtain 
\begin{equation}
J_{\mu}(\omega)= \sigma_{\mu\nu}(\omega)E_{\nu},
\end{equation}
where the conductivity is written as
\begin{eqnarray}
\sigma_{\mu\nu}(\omega)&=& \frac{i}{\hbar}\int^{\infty}_0 dt{\rm e}^{i\omega t}
{\rm Tr}{\rm e}^{-iH_0t/\hbar}[a_{\nu},\rho_0]{\rm e}^{iH_0t/\hbar}J_{\mu}\nonumber\\
&=& \int^{\infty}_0 dt{\rm e}^{i\omega t}\phi_{\mu\nu}(t).
\end{eqnarray}
Here we have defined
\begin{equation}
\phi_{\mu\nu}(t)= \frac{i}{\hbar}{\rm Tr}[a_{\nu},\rho_0]{\rm e}^{iH_0t/\hbar}
J_{\mu}{\rm e}^{-iH_0t/\hbar}.
\end{equation}
Due to the relation
\begin{equation}
[a_{\nu},\rho_0]=-i\hbar\rho_0 \int^{\beta}_0 d\lambda\dot{a}_{\nu}(-i\hbar\lambda),
\end{equation}
we obtain
\begin{eqnarray}
\phi_{\mu\nu}(t)&&= \int^{\beta}_0d\lambda{\rm Tr}\rho_0\dot{a}_{\nu}(-i\hbar\lambda)
J_{\mu}(t)\nonumber\\
&&= \int^{\beta}_0 d\lambda\langle J_{\nu}(-i\hbar\lambda)J_{\mu}(t)\rangle.
\end{eqnarray}
Since the time-derivative of $\phi$ is written as
\begin{equation}
\dot{\phi}_{\mu\nu}= -\frac{i}{\hbar}\langle J_{\mu}(t)J_{\nu}(0)-J_{\nu}(0)J_{\mu}(t)\rangle,
\end{equation}
the conductivity is given by
\begin{eqnarray}
\sigma_{\mu\nu}(\omega)&=& \int^{\infty}_0{\rm e}^{i\omega t}\phi_{\mu\nu}(t)
= -\int^{\infty}_0dt\frac{{\rm e}^{i(\omega+i\delta)t}-1}{i(\omega+i\delta)}
\dot{\phi}_{\mu\nu}\nonumber\\
&=&-\int^{\infty}_{-\infty}dt \frac{{\rm e}^{i(\omega+i\delta)t}-1}{i(\omega+i\delta)}
Q^R_{\mu\nu}(t)\nonumber\\
&=&-\frac{1}{i(\omega+i\delta)}[Q^R_{\mu\nu}(\omega)-Q^R_{\mu\nu}(0)],
\end{eqnarray}
where
\begin{equation}
Q^R_{\mu\nu}(t)= -\frac{i}{\hbar}\theta(t)\langle[J_{\mu}(t),J_{\nu}(0)]\rangle,
\end{equation}
and $Q^R_{\mu\nu}(\omega)$ is its Fourier transform.  $Q^R_{\mu\nu}(\omega)$ is
evaluated from the analytic continuation of the thermal Green's function:
\begin{equation}
Q^R_{\mu\nu}(\omega)= Q_{\mu\nu}(i\omega_n\rightarrow \hbar\omega+i\delta),
\end{equation}
\begin{equation}
Q_{\mu\nu}(\tau)=-\langle{\rm T}J_{\mu}(\tau)J_{\nu}(0)\rangle
=\frac{1}{\beta}\sum_{\omega_n}Q_{\mu\nu}(i\omega_n){\rm e}^{-i\omega_n\tau}.
\end{equation}
From these equations we can derive the sum rule for $\sigma_{\mu\nu}$.
Let us define the Fourier transform of $\phi_{\mu\nu}(t)$ as
\begin{equation}
\phi_{\mu\nu}(\omega)= \int^{\infty}_{-\infty}dt\phi_{\mu\nu}(t){\rm e}^{i\omega t},
\end{equation}
then $\sigma_{\mu\nu}(\omega)$ is written as
\begin{equation}
\sigma_{\mu\nu}(\omega)= \frac{i}{2\pi}\int^{\infty}_{-\infty}d\omega'
\phi_{\mu\nu}(\omega')\frac{1}{\omega-\omega'+i\delta}.
\end{equation}
Hence the following formulae are followed:
\begin{equation}
\int^{\infty}_{-\infty}d\omega \sigma_{\mu\nu}(\omega)= \pi\phi_{\mu\nu}(t=0),
\end{equation}
\begin{equation}
\lim_{\omega\rightarrow 0}\omega\sigma_{\mu\nu}(\omega)= i\phi_{\mu\nu}(t=0).
\end{equation}
Since $v_{i\mu}=(1/m)(p_{i\mu}-(e/c)A_{\mu})$ and $J_{\mu}=e\sum_i(v_i)_{\mu}$, 
we obtain
\begin{eqnarray}
\phi_{\mu\nu}(t=0)&=&\frac{i}{\hbar}{\rm Tr}[a_{\nu},\rho_0]J_{\mu}\nonumber\\
&=& 
\frac{i}{\hbar}{\rm Tr}(\rho_0[J_{\mu},\sum_jex_{j\nu}])\nonumber\\
&=&\delta_{\mu\nu}\frac{Ne^2}{m}.
\label{phi0}
\end{eqnarray}
Then the sum rule is written as
\begin{equation}
\int^{\infty}_0 d\omega {\rm Re}\sigma(\omega)= \frac{\pi}{2}\frac{Ne^2}{m}
\end{equation}
In this derivation
the translational invariance of the potential term is important since we used
the relation $\dot{x}_{i\mu}=v_{i\mu}=(1/m)(p_{i\mu}-(e/c)A_{\mu})$.
In the Drude formula
\begin{equation}
\sigma(\omega)= \frac{Ne^2}{m}\frac{\tau}{1-i\omega\tau},
\end{equation}
the sum rule is clearly satisfied.

Please note that the above formulas are derived for the uniform external fields.
An extension to the spatially oscillating fields is performed in a straightforward
way.
Here we set $\hbar=1$.
Let us consider the spatially varying applied fields:
\begin{equation}
H'(t)= -\frac{1}{c}\int d{\bf r}\xi_{\mu}({\bf r})E_{\mu}({\bf r},t),
\end{equation}
or
\begin{equation}
H'(t)= -\frac{1}{c}\int d{\bf r}j_{\mu}({\bf r})A_{\mu}({\bf r},t),
\end{equation}
where we assume
\begin{equation}
E_{\mu}({\bf r},t)= e_{\mu}{\rm e}^{i({\bf q}\cdot{\bf r}-\omega t)},
\end{equation}
and $A_{\mu}({\bf r},t)$ is the vector potential satisfying 
$E_{\mu}({\bf r},t)=(-1/c)\dot{A}_{\mu}({\bf r},t)=(i\omega/c)A_{\mu}({\bf r},t)$
and ${\rm div}{\bf A}=0$.
We set $\xi_{\mu}({\bf r},t)=e\sum_i x_{i\mu}\delta({\bf r}-{\bf r}_i)$ and 
$j_{\mu}$ is the current operator given by 
\begin{equation}
j_{\mu}({\bf r})= \frac{1}{2m}\sum_ie[{\bf p}_i\delta({\bf r}-{\bf r}_i)+
\delta({\bf r}-{\bf r}_i){\bf p}_i]_{\mu}.
\end{equation} 
The conductivity is defined as the coefficient of the linear response of the current
to applied fields:
\begin{eqnarray}
J_{\mu}({\bf r},t)&=& \sigma_{\mu\nu}({\bf q},\omega)E_{\nu}({\bf r},t)\nonumber\\
&=& \int d{\bf r}'\int^t_{-\infty}dt' \sigma_{\mu\nu}({\bf r}-{\bf r}';t-t')
E_{\nu}({\bf r}',t').
\end{eqnarray}
The expectation value of $j_{\mu}({\bf r})$ to the first order in $H'$ is evaluated as
\begin{eqnarray}
\langle j_{\mu}({\bf r},t)\rangle &=&{\rm Tr}\rho'j_{\mu}({\bf r})\nonumber\\
&=& i\int^{\infty}_0 dt'\int d{\bf r}{\rm Tr}{\rm e}^{-iH_0t'}
[\xi_{\nu}({\bf r}),\rho_0]{\rm e}^{iH_0t'}e_{\nu}{\rm e}^{i{\bf q}\cdot{\bf r}}
{\rm e}^{-i\omega (t-t')}j_{\mu}({\bf r})\nonumber\\
&=& \frac{i}{\hbar}\int^{\infty}_0 dt'{\rm e}^{i\omega t'}{\rm Tr}{\rm e}^{-iH_0t'}
[\xi_{\nu}({\bf q}),\rho_0]{\rm e}^{iH_0t'}e_{\nu}
{\rm e}^{-i\omega t}j_{\mu}({\bf r}),
\end{eqnarray}
where $\xi_{\mu}({\bf q})$ is the Fourier transform of $\xi_{\mu}({\bf r})$.
We follow the same procedure as for the uniform external fields and note that
$\dot{\xi}_{\mu}({\bf r})=J_{\mu}({\bf r})$, then we have
\begin{equation}
\phi_{\mu\nu}({\bf q},t)= \int^{\beta}_0d\lambda\langle j_{\nu}({\bf q},-i\hbar\lambda)
j_{\mu}({\bf r},t)\rangle{\rm e}^{-i{\bf q}\cdot{\bf r}}.
\end{equation}
Here we neglect the ${\bf A}$-term in the current $J_{\mu}$ since this is the 
higher-order term in external
fields and we take a spatial average to write
\begin{equation}
j_{\mu}({\bf q})= \int d{\bf r}j_{\mu}({\bf r}){\rm e}^{i{\bf q}\cdot{\bf r}}
=\frac{e}{2m}\sum_i ({\bf p}_i{\rm e}^{i{\bf q}\cdot{\bf r}_i}
+{\rm e}^{i{\bf q}\cdot{\bf r}_i}{\bf p}_i)_{\mu},
\end{equation}
\begin{equation}
\phi_{\mu\nu}({\bf q},t)= \int^{\beta}_0 d\lambda\langle j_{\nu}({\bf q},
-i\hbar\lambda)j_{\mu}({\bf q},t)\rangle,
\end{equation}
\begin{equation}
\sigma_{\mu\nu}({\bf q},\omega)= \int^{\infty}_0 dt{\rm e}^{i\omega t}
\phi_{\mu\nu}({\bf q},t).
\end{equation}
Now let us define the retarded
response function:
\begin{equation}
K_{\mu\nu}({\bf q},t-t')= -i\theta(t-t')\langle[
j_{\mu}^{\dag}({\bf q},t),j_{\nu}({\bf q},t')]\rangle,
\end{equation}
and its Fourier transform given by
\begin{equation}
K_{\mu\nu}({\bf q},\omega)= \int^{\infty}_{-\infty}dt{\rm e}^{i\omega t}
K_{\mu\nu}({\bf q},t).
\end{equation}
The formula for the optical conductivity is followed as
\begin{equation}
\sigma_{\mu\nu}({\bf q},\omega)= \frac{i}{\omega+i\delta}[
K_{\mu\nu}({\bf q},\omega)-K_{\mu\nu}({\bf q},0)].
\label{sformula}
\end{equation}
The current response function is written in the form,
\begin{equation}
K_{\mu\nu}({\bf q},\omega+i\delta)=-\sum_{nm}\frac{{\rm e}^{-\beta E_n}}{Z}[
\frac{\langle n|j_{\nu}({\bf q})|m\rangle\langle m|j^{\dag}_{\mu}({\bf q})|n\rangle}
{E_m-E_n+\omega+i\delta}
-\frac{\langle n|j^{\dag}_{\mu}({\bf q})|m\rangle\langle m|j_{\nu}({\bf q})|n\rangle}
{E_n-E_m+\omega+i\delta}],
\end{equation}
where $|n\rangle$ denotes a complete set of exact eigenstates of $H_0$ with
eigenvalues $E_n$ and $Z$ is the partition function $Z=\sum_n{\rm e}^{-\beta E_n}$.
The imaginary part of $K_{\mu\nu}$ is given by
\begin{equation}
{\rm Im}K_{\mu\nu}({\bf q},\omega+i\delta)=-\pi(1-{\rm e}^{-\beta\omega})
\sum_{nm}\frac{{\rm e}^{-\beta E_n}}{Z}\delta(E_m-E_n-\omega)
\langle n|j^{\dag}_{\mu}({\bf q})|m\rangle\langle m|j_{\nu}({\bf q})|n\rangle.
\end{equation}
The retarded Green's function
$K_{\mu\nu}({\bf q},\omega)$ is evaluated from the thermal Green's function
through the analytic continuation:
\begin{equation}
K_{\mu\nu}({\bf q},\omega)= K_{\mu\nu}({\bf q},i\omega_n=\omega+i\delta),
\end{equation}
where
\begin{equation}
K_{\mu\nu}({\bf q},i\omega_n)= \int^{\beta}_0 d\tau {\rm e}^{i\omega_n\tau}
K_{\mu\nu}({\bf q},\tau),
\end{equation}
\begin{equation}
K_{\mu\nu}({\bf q},\tau)= -\langle Tj^{\dag}_{\mu}({\bf q},\tau)j_{\nu}({\bf q},0)
\rangle.
\end{equation}

In order to calculate the conductivity for the uniform external fields, we take
the limit ${\bf q}\rightarrow 0$.  For the direct-current conductivity, we must
take the limit ${\bf q}\rightarrow 0$ first before $\omega \rightarrow 0$.
The current operator is given by
\begin{equation}
{\bf j}({\bf q})= \frac{e}{m}\sum_{{\bf p}\sigma}({\bf p}+\frac{1}{2}{\bf q})
c^{\dag}_{{\bf p}+{\bf q}\sigma}c_{{\bf p}\sigma}
=\frac{e}{m}\sum_{{\bf p}\sigma}{\bf p}c^{\dag}_{{\bf p}+{\bf q}/2,\sigma}
c_{{\bf p}-{\bf q}/2,\sigma}.
\end{equation}
In the limit ${\bf q}\rightarrow 0$ $K_{\mu\nu}({\bf q},\omega=0)$ is evaluated as
\begin{eqnarray}
K_{\mu\nu}(0,0)&=& \frac{2e^2}{m^2}\frac{1}{\beta}\sum_n\sum_{{\bf k}}k_{\mu}k_{\nu}
G_0({\bf k},i\epsilon_n)^2\nonumber\\
&=& \frac{2e^2}{m^2}\frac{1}{\beta}\sum_n\sum_{\bf k}k_{\mu}\frac{m}{2}
\frac{\partial}{\partial k_{\nu}}G_0({\bf k},i\epsilon_n)
= \frac{2e^2}{m^2}\frac{m}{2}\frac{1}{\beta}\sum_n\sum_{\bf k}(-\delta_{\mu\nu})
G_0({\bf k},i\epsilon_n)\nonumber\\
&=& -\frac{Ne^2}{m}\delta_{\mu\nu},
\end{eqnarray}
where $G_0({\bf k},i\epsilon_n)$ is the Green's function for the non-interacting
electrons: $G_0({\bf k},i\epsilon_n)=(i\epsilon_n-\epsilon_{{\bf k}})^{-1}$ for
$\epsilon_n=(2n+1)\pi/\beta$.
Then the uniform conductivity is given by the formula
\begin{equation}
\sigma_{\mu\nu}(\omega)= \frac{i}{\omega+i\delta}[K_{\mu\nu}({\bf q}=0,\omega+i\delta)
+\frac{Ne^2}{m}\delta_{\mu\nu}].
\end{equation}
The current response function $K_{\mu\nu}$ for ${\bf q}\neq 0$ in the
non-intracting system is
\begin{equation}
K_{\mu\nu}({\bf q},i\omega_n)= -\frac{2e^2}{m^2}\sum_{\bf p}p_{\mu}p_{\nu}
\frac{f(\epsilon_{{\bf p}-{\bf q}/2})-f(\epsilon_{{\bf p}+{\bf q}/2})}
{-i\omega_n+\epsilon_{{\bf p}-{\bf q}/2}-\epsilon_{{\bf p}+{\bf q}/2}}.
\end{equation}
Since $K_{\mu\nu}({\bf q},\omega)$ vanishes in the limit ${\bf q}\rightarrow 0$ 
for finite $\omega\neq 0$, we have the sum rule
\begin{equation}
\int^{\infty}_0d\omega {\rm Re}\sigma_{\mu\nu}(\omega)= \frac{\pi}{2}
\frac{Ne^2}{m}\delta_{\mu\nu},
\end{equation} 
and the Drude weight
\begin{equation}
D= \frac{\pi Ne^2}{m}
\end{equation}
as the coefficient of the delta function: 
Re$\sigma_{\mu\nu}(\omega)=D\delta(\omega)\delta_{\mu\nu}$.
The sum rule for ${\bf q}\neq 0$ is derived similarly.  The commutator
$[j^{\dag}_{\mu},\xi_{\nu}]$ as in eq.(\ref{phi0}) leads to
\begin{equation}
\int^{\infty}_{-\infty}d\omega {\rm Re}\sigma_{\mu\nu}({\bf q},\omega)
=\pi{\rm Re}\phi_{\mu\nu}({\bf q},t=0)=\frac{\pi Ne^2}{m}\delta_{\mu\nu}.
\end{equation}

\subsection{Mattis-Bardeen Formula}
The infrared absorption formula in BCS model was first derived by 
Mattis-Bardeen.\cite{mat58}
We use the standard notations for the Green's functions;
\begin{equation}
G({\bf p},\tau)=-\langle c_{{\bf p}\sigma}(\tau)c^{\dag}_{{\bf p}\sigma}(0)
\rangle,
\end{equation}
\begin{equation}
F({\bf p},\tau)= \langle c_{-{\bf p}\downarrow}(\tau)c_{{\bf p}\uparrow}(0)
\rangle,
\end{equation}
\begin{equation}
F^{\dag}({\bf p},\tau)= \langle c^{\dag}_{{\bf p}\uparrow}(\tau)
c^{\dag}_{-{\bf p}\downarrow}(0)\rangle,
\end{equation}
and their Fourier transforms given by
\begin{equation}
G({\bf p},i\epsilon_n)= \frac{u^2_{{\bf p}}}{i\epsilon_n-E_{{\bf p}}}+
\frac{v^2_{{\bf p}}}{i\epsilon_n+E_{{\bf p}}},
\end{equation}
\begin{equation}
F({\bf p},i\epsilon_n)= F^{\dag}({\bf p},i\epsilon_n)= -u_{{\bf p}}v_{{\bf p}}
(\frac{1}{i\epsilon_n-E_{{\bf p}}}-\frac{1}{i\epsilon_n+E_{{\bf p}}}),
\end{equation}
where $E_{{\bf p}}=\sqrt{\xi_{{\bf p}}^2+\Delta^2_{{\bf p}}}$,
$u^2_{{\bf p}}=(1/2)(1+\xi_{{\bf p}}/E_{{\bf p}})$ and
$v^2_{{\bf p}}=1-u^2_{{\bf p}}$ for $\xi_{{\bf p}}=\epsilon_{{\bf p}}-\mu$.
We evaluate the current response function written as
\begin{eqnarray}
K_{\mu\nu}({\bf q},i\omega_{\ell})&=& \frac{2e^2}{m^2}\sum_{{\bf p}}p_{\mu}p_{\nu}
\frac{1}{\beta}\sum_n [G({\bf p}-{\bf q}/2,i\epsilon_n)G({\bf p}+{\bf q}/2,
i\epsilon_n-i\omega_{\ell})\nonumber\\
&+& F({\bf p}-{\bf q}/2,i\epsilon_n)
F^{\dag}({\bf p}+{\bf q}/2,i\epsilon_n+i\omega_{\ell})].
\end{eqnarray}
We set $i\omega_{\ell}\rightarrow -i\omega_{\ell}$ in the first term, then using
the symmetry ${\bf q}\leftrightarrow -{\bf q}$ we have
\begin{eqnarray}
K_{\mu\nu}({\bf q},i\omega_{\ell})&=& \frac{2e^2}{m^2}\sum_{{\bf p}}p_{\mu}p_{\nu}
[ (f(E_{{\bf p}_+})-f(E_{{\bf p}_-}))\frac{1}{2}\left(1+
\frac{\xi_{{\bf p}_+}\xi_{{\bf p}_-}}{E_{{\bf p}_+}E_{{\bf p}_-}}
+\frac{ \Delta_{{\bf p}_+}\Delta_{{\bf p}_-} }{ E_{{\bf p}_+}E_{{\bf p}_-} }\right)
\frac{1}{i\omega_{\ell}+E_{{\bf p}_+}-E_{{\bf p}_-} }\nonumber\\
&+& (f(E_{{\bf p}_+})+f(E_{{\bf p}_-})-1)\frac{1}{4}\left( (1+
\frac{\xi_{{\bf p}_+}}{E_{{\bf p}_+}})(1-\frac{\xi_{{\bf p}_-}}
{E_{{\bf p}_-}})-\frac{\Delta_{{\bf p}_+}\Delta_{{\bf p}_-}}
{E_{{\bf p}_+}E_{{\bf p}_+}}\right)
\frac{1}{i\omega_{\ell}+E_{{\bf p}_+}+E_{{\bf p}_-} }\nonumber\\
&+& (1-f(E_{{\bf p}_+})-f(E_{{\bf p}_-}))\frac{1}{4}\{(1-
\frac{\xi_{{\bf p}_+}}{E_{{\bf p}_+}})(1+\frac{\xi_{{\bf p}_-}}
{E_{{\bf p}_-}})-\frac{\Delta_{{\bf p}_+}\Delta_{{\bf p}_-}}
{E_{{\bf p}_+}E_{{\bf p}_+}}\}
\frac{1}{i\omega_{\ell}-E_{{\bf p}_+}-E_{{\bf p}_-} }],\nonumber\\
\end{eqnarray}
where ${\bf p}_+={\bf p}+{\bf q}/2$ and ${\bf p}_-={\bf p}-{\bf q}/2$.
Here we consider (i) dirty superconductors  or (ii) thin films satisfying
$d\ll\xi\approx v_F/\Delta$ for the film thickness $d$ and the coherence
length $\xi$.
In these cases we can regard ${\bf p}_+$ and ${\bf p}_-$ as independent
variables.  In the case (ii), since $qv_F\gg\Delta\sim\omega$ we can do 
Abrikosov's replacement\cite{abr88},
\begin{equation}
\sum_{{\bf p}}\rightarrow N(0)\frac{1}{4qv_F}\int d\xi_{{\bf p}_+} d\xi_{{\bf p}_-}.
\end{equation}
Then in the isotropic case we obtain the Mattis-Bardeen formulae for 
$\sigma_1(\omega)={\rm Re}\sigma(\omega)=-(1/\omega){\rm Im}K(\omega+i\delta)$ and
$\sigma_2(\omega)={\rm Im}\sigma(\omega)=(1/\omega){\rm Re}K(\omega+i\delta)$:
\begin{eqnarray}
\frac{\sigma_{1s}}{\sigma_{1n}}(\omega)&=& \frac{1}{\omega}
\int^{\omega-\Delta}_{\Delta}dEN(E)N(\omega-E)(1-2f(\omega+E))\left( 1-
\frac{\Delta^2}{E(\omega-E)}\right)\theta(\omega-2\Delta)\nonumber\\
&+& 2\frac{1}{\omega}\int^{\infty}_0dE(f(E)-f(\omega+E))\left( 1+
\frac{\Delta^2}{E(\omega+E)}\right), 
\end{eqnarray}
\begin{equation}
\frac{\sigma_{2s}}{\sigma_{1n}}(\omega)= \frac{1}{\omega} 
\int^{\Delta}_{max(-\Delta,\Delta-\omega)}dE(1-2f(\omega+E))
\frac{E(E+\omega)+\Delta^2}{\sqrt{\Delta^2-E^2}\sqrt{(E+\omega)^2-\Delta^2}},
\end{equation}
where $\sigma_{1n}$ is the real part of the conductivity for normal state.
In the above formula for $\sigma_2$, the integral is calculated as follows.
\begin{eqnarray}
I&\equiv& \frac{1}{4}\int^{\infty}_{-\infty} d\xi_p d\xi_{p'}
(1-\frac{\xi_p\xi_{p'}+\Delta^2}{E_pE_{p'}})
{\rm P}\frac{1}{\omega-E_p-E_{p'}}\nonumber\\
&=& \int^{\infty}_{\Delta}dEdE'N(E)N(E')(1-\frac{\Delta^2}{EE'})
{\rm P}\frac{1}{\omega-E_p-E_{p'}}\nonumber\\
&=& \int^{\infty}_{-\infty}d\epsilon \int^{\infty}_{\Delta}dEdE'N(E)N(E')
(1-\frac{\Delta^2}{EE'}){\rm P}\frac{1}{\omega-\epsilon}\delta(\epsilon+E-E')
\nonumber\\
&=& \int^{\infty}_{\Delta-E}d\epsilon \int^{\infty}_{\Delta}dE N(E)N(E+\epsilon)
(1-\frac{\Delta^2}{E(E+\epsilon)}){\rm P}\frac{1}{\omega-\epsilon}\nonumber\\
&=& {\rm Re}\int^{\infty}_{-\infty}d\epsilon\int^{\infty}_{\Delta}dEN(E)N(E+\epsilon)
(1-\frac{\Delta^2}{E(E+\epsilon)})\frac{-1}{2}(\frac{1}{\epsilon+\omega+i\delta}
+\frac{1}{\epsilon+\omega-i\delta})\nonumber\\
&=& \pi {\rm Im}\int^{\infty}_{\Delta}dEN(E)N(E-\omega)
(1-\frac{\Delta^2}{E(E-\omega)})\nonumber\\
&=& -\pi\int^{\omega+\Delta}_{max(\omega-\Delta,\Delta)}dE
\frac{E(\omega-E)-\Delta^2}{\sqrt{E^2-\Delta^2}\sqrt{\Delta^2-(\omega-E)^2}}\nonumber\\
&=& \pi \int^{\Delta}_{max(\Delta-\omega,-\Delta)}dE
\frac{E(\omega+E)+\Delta^2}{\sqrt{\Delta^2-E^2}\sqrt{(\omega-E)^2-\Delta^2}}.
\end{eqnarray}
At $T=0$, the integral has the form for complete elliptic integrals by changing
variables of integration to $x$ where $x=(2E-\omega)/(\omega-2\Delta)$ for 
$\sigma_{1s}$;  
\begin{eqnarray}
\frac{\sigma_{1s}}{\sigma_{1n}}(\omega)&=& \frac{1}{\omega}
\int_{\Delta}^{\omega-\Delta} dE\frac{E(\omega-E)-\Delta^2}
{\sqrt{E^2-\Delta^2}\sqrt{(\omega-E)^2-\Delta^2}}\nonumber\\
&=& \frac{\omega-2\Delta}{\omega}\int_0^1 dx \frac{1-kx^2}
{\sqrt{(1-x^2)(1-k^2x^2)}},
\end{eqnarray}
where $k=(\omega-2\Delta)/(\omega+2\Delta)$ and $\omega >2\Delta$.
We then obtain the Mattis-Bardeen formula:
\begin{equation}
\frac{\sigma_{1s}}{\sigma_{1n}}(\omega)=\left(\left(1+\frac{2\Delta}{\omega}\right)
E(k)-\frac{4\Delta}{\omega} K(k)\right)\theta(\omega-2\Delta),
\end{equation}
where   
\begin{equation}
E(k)= \int_0^1\sqrt{\frac{1-k^2x^2}{1-x^2}} dx,~~
K(k)= \int_0^1\frac{1}{\sqrt{(1-x^2)(1-k^2x^2)}}dx,
\end{equation}
are complete elliptic integrals.
An expression for $\sigma_{2s}$ valid
for all $\omega$ is
\begin{equation}
\frac{\sigma_{2s}}{\sigma_{1n}}(\omega)= \left(\frac{\Delta}{\omega}+\frac{1}{2}
\right)E(k')+\left(\frac{\Delta}{\omega}-\frac{1}{2}\right)K(k'),
\end{equation}
where $k'=(1-k^2)^{1/2}$.

In evaluations of $K_{\mu\nu}$,  the matrix notations are also employed 
for Green's functions in superconductors,
\begin{equation}
\hat{G}({\bf k},i\epsilon_n)= \frac{i\tilde{\epsilon_n}\tau_0+\xi_{{\bf k}}\tau_3
+\tilde{\Delta}_{{\bf k}}\tau_1}{\tilde{\epsilon_n}^2+\xi_{{\bf k}}^2
+\tilde{\Delta_{{\bf k}}}^2},
\end{equation}
where $\tau_i$ ($i=0,1,3$) are Pauli matrices.  $\epsilon_n$ and $\Delta_{{\bf k}}$
are generalized to include the self-energy in the form:
$\tilde{\epsilon_n}=\epsilon_n-\Sigma_0(\epsilon_n)$ and
$\tilde{\Delta_{{\bf k}}}= \Delta_{{\bf k}}+\Sigma_1(\epsilon_n)$. 
The response function $K_{\mu\nu}$ is written as neglecting pair vertex corrections
\begin{equation}
K_{\mu\nu}({\bf q},i\omega_m)= \frac{e^2}{m^2}\sum_{{\bf p}}p_{\mu}p_{\nu}
\frac{1}{\beta}\sum_n {\rm Tr}\hat{G}({\bf p}_+,i\epsilon_n+i\omega_m)
\hat{G}({\bf p}_-,i\epsilon_n).
\label{kmn}
\end{equation}
One can reproduce the Mattis-Bardeen formulae from this expression in a 
straightforward way.

\subsection{Optical Conductivity in London Superconductor}
In this section, we investigate the superconductor in the London limit:
$\xi\ll \lambda$ for the coherence length $\xi$ and the penetration depth
$\lambda$.  Since $q\xi\ll 1$ holds, we must consider the limit 
$q\rightarrow 0$ for the response function in 
eq.(\ref{kmn})\cite{ska64,hir89,hir92,hir94,dor00,dor03}:
\begin{equation}
K_{\mu\nu}({\bf q},i\omega_m)= \frac{e^2k_F^2}{m^2}2\sum_{{\bf k}}\hat{k}_{\mu}
\hat{k}_{\nu}
\frac{1}{\beta}\sum_n 
\frac{-\tilde{\epsilon}_n(\tilde{\epsilon}_n+\omega_m)+\xi_{{\bf k}_-}\xi_{{\bf k}_+}
+\tilde{\Delta}_{{\bf k}_-}\tilde{\Delta}_{{\bf k}_+} }
{(\tilde{\epsilon}_n^2+\xi^2_{{\bf k}_-}+\tilde{\Delta}^2_{{\bf k}_-})
((\tilde{\epsilon}_n+\omega_m)^2+\xi^2_{{\bf k}_+}+\tilde{\Delta}^2_{{\bf k}_+})}.
\end{equation}
We use the notation $\tilde{z}=i\tilde{\epsilon}_n=i\epsilon_n-i\Sigma_0(i\epsilon_n)=z-i\Sigma_0(z)$.  Then we have
\begin{equation}
K_{\mu\nu}({\bf q}=0,i\omega_m)= \frac{e^2k_F^2}{m^2}2\sum_{{\bf k}}\hat{k}_{\mu}
\hat{k}_{\nu}\frac{1}{2\pi i}\int_C dzf(z)
\frac{\tilde{z}(\tilde{z}+i\omega_m)+\xi^2_{{\bf k}}
+\tilde{\Delta}_{{\bf k}}\tilde{\Delta}_{{\bf k}_+} }
{(\xi^2_{{\bf k}}+\tilde{\Delta}^2_{{\bf k}}-\tilde{z}^2)(\xi^2_{{\bf k}}
+\tilde{\Delta}^2_{{\bf k}_+}-(\tilde{z}+i\omega_m)^2)},
\end{equation}
where $\tilde{\Delta}^2_{{\bf k}_+}=\Delta_{{\bf }}+\Sigma_1(z+i\omega)$ and $C$ is
the contour surrounding the poles of $f(z)$ in the clockwise direction.
We set $z_+=z+i\omega_m$, and write the integrand in the form,
\begin{eqnarray}
\frac{\tilde{z}\tilde{z}_++\xi^2_{{\bf k}}+\tilde{\Delta}^2_{{\bf k}}
\tilde{\Delta}^2_{{\bf k}_+}}
{(\xi^2_{{\bf k}}+\tilde{\Delta}^2_{{\bf k}}-\tilde{z}^2)(\xi^2_{{\bf k}}
+\tilde{\Delta}^2_{{\bf k}_+}-(\tilde{z}+i\omega_m)^2)}
&=& \frac{1}{\xi^2_{{\bf k}}+\tilde{\Delta}^2_{{\bf k}}-\tilde{z}^2}\nonumber\\
&+& \frac{\tilde{z}\tilde{z}_+ +\tilde{\Delta}^2_{{\bf k}}\tilde{\Delta}^2_{{\bf k}_+}
+\tilde{z}^2_+ -\tilde{\Delta}^2_{{\bf k}_+} }
{(\xi^2_{{\bf k}}+\tilde{\Delta}^2_{{\bf k}}-\tilde{z}^2)(\xi^2_{{\bf k}}
+\tilde{\Delta}^2_{{\bf k}_+}-(\tilde{z}+i\omega_m)^2)}.\nonumber\\
\label{ker1}
\end{eqnarray}
Then the momentum summations are performed in the following way:
\begin{eqnarray}
\sum_{{\bf k}}\frac{1}{2\pi i}\int_C dzf(z)\frac{1}
{\xi^2_{{\bf k}}+\tilde{\Delta}^2_{{\bf k}}-\tilde{z}^2}
&=& N(0)\langle\int_{-\infty}^{\infty}d\xi\frac{1}{2\pi i}\int_C dzf(z)
\frac{z}{(\xi^2+\tilde{\Delta}^2_{{\bf k}}-\tilde{z}^2)^2}
\frac{\partial}{\partial z}(\tilde{\Delta}^2_{{\bf k}}-\tilde{z}^2)\rangle_{\hat{k}}
\nonumber\\
&=& N(0)\langle\frac{1}{2i}\int_C dz[-\frac{\partial}{\partial z}
\frac{zf(z)}{ \sqrt{\tilde{\Delta}^2_{{\bf k}}-\tilde{z}^2} }
+\frac{f(z)}{\sqrt{\tilde{\Delta}^2_{{\bf k}}-\tilde{z}^2}}]\rangle_{\hat{k}},
\end{eqnarray}
where $\langle\cdots\rangle_{\hat{k}}$ denotes the average over the Fermi
surface.  In the last equality, the first term gives only a constant 
contribution.  The second term in eq.(\ref{ker1}) gives
\begin{eqnarray}
\int^{\infty}_{-\infty}d\xi_{{\bf k}}\frac{1}
{(\xi^2_{{\bf k}}+\tilde{\Delta}^2_{{\bf k}}-\tilde{z}^2)(\xi^2_{{\bf k}}
+\tilde{\Delta}^2_{{\bf k}_+}-(\tilde{z}+i\omega_m)^2)}
&=& \pi i\frac{1}{ \sqrt{\tilde{z}^2-\tilde{\Delta}^2_{{\bf k}}}
\sqrt{\tilde{z}_+^2-\tilde{\Delta}^2_{{\bf k}_+}} }\nonumber\\
&\times& \frac{-1}{ \sqrt{\tilde{z}^2-\tilde{\Delta}^2_{{\bf k}}}
+\sqrt{\tilde{z}_+^2-\tilde{\Delta}^2_{{\bf k}_+}} }.
\end{eqnarray}
Then the current response function is written as
\begin{eqnarray}
K_{\mu\nu}({\bf q}=0,i\omega_m)&=& \frac{e^2k_F^2}{m^2}N(0)\langle k_{\mu}
k_{\nu}\int_C dzf(z)
\frac{1}{ \sqrt{\tilde{z}^2-\tilde{\Delta}^2_{{\bf k}}}
+\sqrt{\tilde{z}_+^2-\tilde{\Delta}^2_{{\bf k}_+}} }\nonumber\\
&\times& \left(1-
\frac{\tilde{z}\tilde{z}_+ +\tilde{\Delta}_{{\bf k}}\tilde{\Delta}_{{\bf k}_+}}
{ \sqrt{\tilde{z}^2-\tilde{\Delta}^2_{{\bf k}}}
\sqrt{\tilde{z}_+^2-\tilde{\Delta}^2_{{\bf k}_+}} }\right)\rangle_{\hat{k}}.
\end{eqnarray}

Let us consider the limit of weak impurity scattering, and write the
renormalized frequency $\tilde{z}$ and the superconducting gap 
$\tilde{\Delta}_{{\bf k}}$ in the following forms, respectively:
\begin{equation}
\tilde{z}= z+i\Gamma_1\langle\frac{\tilde{z}}
{ \sqrt{\tilde{z}^2-\tilde{\Delta}^2_{{\bf k}}} }\rangle_{\hat{k}},
\end{equation}
\begin{equation}
\tilde{\Delta}_{{\bf k}}= \Delta_{{\bf k}}+i\Gamma_2\langle
\frac{ \tilde{\Delta}_{{\bf k}'} }
{ \sqrt{\tilde{z}^2-\tilde{\Delta}^2_{{\bf k}'}} }\rangle_{\hat{k}'}.
\end{equation}
For the isotropic superconducting gap, we set $u=\tilde{z}/\tilde{\Delta}$
and $v=u\Delta$; then we have
\begin{equation}
v(z)= z+i\Gamma \frac{u}{\sqrt{u^2-1}}=z+i\Gamma \frac{v}{\sqrt{v^2-\Delta^2}},
\end{equation}
where $\Gamma=\Gamma_1-\Gamma_2$.  The relation
$\sqrt{\tilde{z}^2-\tilde{\Delta}^2}=\sqrt{v^2-\Delta^2}+i\Gamma_2$ is also
followed.  The density of states is given by
\begin{equation}
N_s(z)= N(0){\rm Re}\frac{u}{\sqrt{u^2-1}}.
\end{equation}
In the limit as $\Gamma\rightarrow 0$, this reduces to
\begin{equation}
N_s(\omega)= N(0)\frac{|\omega|}{\sqrt{\omega^2-\Delta^2}}~~{\rm for}~~
|\omega|>\Delta,
\label{doseq}
\end{equation}  
and $N_s(\omega)=0$ otherwise.
For the anisotropic case satisfying 
$\langle \Delta_{{\bf k}}\rangle_{\hat{k}}=0$, $v(z)$ satisfies
\begin{equation}
v(z)= \tilde{z}= z+i\Gamma\langle\frac{v(z)}
{ \sqrt{v(z)^2-\Delta^2_{{\bf k}}} }\rangle_{\hat{k}},
\label{vzeq}
\end{equation}
for $\Gamma=\Gamma_1$ ($\Gamma_2=0$).  Let us examine this case in more detail;
$K_{\mu\nu}$ is given by(where we use the notation $v_+=v(z+i\omega_m)$),
\begin{eqnarray}
K_{\mu\nu}({\bf q}=0,i\omega_m)&=& \frac{e^2k_F^2}{m^2}N(0)\langle k_{\mu}
k_{\nu} \int_C dzf(z) \frac{1}
{ \sqrt{v^2-\Delta_{{\bf k}}^2}+\sqrt{v_+^2-\Delta_{{\bf k}}^2} }\nonumber\\
&\times& \left(1-
\frac{vv_++\Delta_{{\bf k}}^2}
{ \sqrt{v^2-\Delta_{{\bf k}}^2}\sqrt{v_+^2-\Delta_{{\bf k}}^2} }\right)
\rangle_{\hat{k}}\nonumber\\
&=& -\frac{e^2k_F^2}{m^2}N(0)\langle k_{\mu}k_{\nu}\int_C dzf(z) \frac{1}
{v_+-v}\nonumber\\
&\times& \left( \frac{v_+}{\sqrt{v_+^2-\Delta_{{\bf k}}^2}}-
\frac{v}{\sqrt{v^2-\Delta_{{\bf k}}^2}}\right)\rangle_{\hat{k}}.
\end{eqnarray}
Now $v$ has a cut along the real axis; as the cut is crossed, the continuation is
performed as follows\cite{ska64},
\begin{eqnarray}
v&\rightarrow& v*,\nonumber\\
(v^2-\Delta^2)^{-1/2}&\rightarrow& -[(v^2-\Delta^2)^{-1/2}]^*,\nonumber\\
\frac{v}{\sqrt{v^2-\Delta^2}}&\rightarrow& -\left(\frac{v}{\sqrt{v^2-\Delta^2}}
\right)^*.
\end{eqnarray}
The complex integration is reduced to integrations along Im$z=0$ and Im$z=\omega_m$.
We obtain the expression for the current response function from the
analytic continuation:
\begin{eqnarray}
K({\bf q}=0,\omega+i\delta)&=& \frac{e^2k_F^2}{2m^2}N(0)\int^{\infty}_{-\infty}
d\epsilon [(f(\epsilon)-f(\epsilon+\omega))\nonumber\\  
&\times& \{ \frac{1}{v(\epsilon+\omega)-v(\epsilon)^*}\langle
\frac{v(\epsilon+\omega)}{\sqrt{v(\epsilon+\omega)^2-\Delta_{{\bf k}}^2}}+
\frac{v(\epsilon)^*}{\sqrt{v(\epsilon)^{*2}-\Delta_{{\bf k}}^2}}\rangle_{\hat{k}}
\nonumber\\
&-& \frac{1}{v(\epsilon+\omega)-v(\epsilon)}\langle
\frac{v(\epsilon+\omega)}{\sqrt{v(\epsilon+\omega)^2-\Delta_{{\bf k}}^2}}-
\frac{v(\epsilon)}{\sqrt{v(\epsilon)^2-\Delta_{{\bf k}}^2}}\rangle_{\hat{k}}\}
\nonumber\\
&-& f(\epsilon+\omega)\nonumber\\
&\times& \{ \frac{1}{v(\epsilon+\omega)^*-v(\epsilon)^*}\langle
\frac{v(\epsilon+\omega)^*}{\sqrt{v(\epsilon+\omega)^{*2}-\Delta_{{\bf k}}^2}}-
\frac{v(\epsilon)^*}{\sqrt{v(\epsilon)^{*2}-\Delta_{{\bf k}}^2}}\rangle_{\hat{k}}
\nonumber\\
&+& \frac{1}{v(\epsilon+\omega)-v(\epsilon)}\langle
\frac{v(\epsilon+\omega)}{\sqrt{v(\epsilon+\omega)^2-\Delta_{{\bf k}}^2}}-
\frac{v(\epsilon)}{\sqrt{v(\epsilon)^2-\Delta_{{\bf k}}^2}}\rangle_{\hat{k}}\}],
\label{kfun}
\end{eqnarray}
where the average $K_{xx}+K_{yy}$ is written as $2K$ (in two dimensions) and we use 
$v(\omega-i\delta)= v(\omega+i\delta)^*$.
If we use the relation in eq.(\ref{vzeq}), the expression for ${\rm Im}K$ is
simplified as
\begin{eqnarray}
{\rm Im}K({\bf q}=0,\omega+i\delta)&=&\frac{e^2k_F^2}{2m^2}N(0)\frac{\omega}{2\Gamma}
\int^{\infty}_{-\infty}d\epsilon \left({\rm tanh}\left(\frac{\beta\epsilon}{2}\right)
-{\rm tanh}\left(\frac{\beta(\epsilon+\omega)}{2}\right)\right)\nonumber\\
&\times& {\rm Re}\left( \frac{1}{v(\epsilon+\omega)-v(\epsilon)}-
\frac{1}{v(\epsilon+\omega)-v(\epsilon)^*} \right).
\end{eqnarray}
In the collision less limit $\Gamma\rightarrow 0$, an expansion in terms of $\Gamma$
gives the conductivity,
\begin{eqnarray}
\frac{\sigma_{1s}}{\sigma_{1n}}(\omega)&=&\frac{1}{2\omega}\int^{\infty}_{-\infty}
d\epsilon \left({\rm tanh}\left(\frac{\beta\epsilon}{2}\right)
-{\rm tanh}\left(\frac{\beta(\epsilon+\omega)}{2}\right)\right)
\langle{\rm Re}\frac{|\epsilon+\omega|}{\sqrt{(\epsilon+\omega)^2-\Delta_{{\bf k}}^2}}
\rangle_{\hat{k}}\nonumber\\
&\times& \langle{\rm Re}\frac{|\epsilon|}{\sqrt{\epsilon^2-\Delta_{{\bf k}}^2}}
\rangle_{\hat{k}},
\end{eqnarray}
where 
$\sigma_{1n}=(ne^2\tau)/m/(\omega\tau)^2\approx(ne^2\tau)/m\cdot 1/[(\omega\tau)^2+1]$
with $\tau=1/(2\Gamma)$, and we use the density of states
in eq.(\ref{doseq}) in the limit $\Gamma\rightarrow 0$.

For the $d$-wave symmetric superconducting gap in two dimensions, the average over
the Fermi surface is given by the Elliptic function,
\begin{equation}
\langle \frac{v}{\sqrt{v^2-\Delta^2{\rm cos}^2(2\phi)}}\rangle_{\hat{k}}=
\frac{1}{2\pi}\int^{2\pi}_0 d\phi\frac{v}{\sqrt{v^2-\Delta^2{\rm cos}^2(2\phi)}}
=\frac{2}{\pi}K(\frac{\Delta}{v}),
\end{equation}
for $v>0$ and $\Delta_{{\bf k}}=\Delta{\rm cos}(2\phi)$.
Since the relation $K(1/x)=-ix/\sqrt{1-x^2}\cdot K(1/\sqrt{1-x^2})$ holds for $x>1$,
the renormalized frequency $u(\omega)=v(\omega)/\Delta$ is written as
\begin{eqnarray}
u(\omega)&=& \frac{\omega}{\Delta}+i\frac{\Gamma}{\Delta}\langle
\frac{u}{\sqrt{u^2-{\rm cos}^2(2\phi)}}\rangle_{\hat{k}}\nonumber\\
&=& \frac{\omega}{\Delta}+\frac{\Gamma}{\Delta}\frac{2}{\pi}\frac{u}{\sqrt{1-u^2}}
K(\frac{1}{\sqrt{1-u^2}}).
\end{eqnarray}
We will do a continuation of the elliptic integral to a general complex argument,
the real part of the optical conductivity for the $d$-wave superconductor in the 
London limit is given by\cite{dor00}
\begin{eqnarray}
\sigma_{1s}(\omega)&=& \frac{e^2k_F^2}{2m^2}N(0)\frac{1}{\omega}\int^{\infty}_{-\infty}
d\epsilon\frac{1}{2} 
\left({\rm tanh}\left(\frac{\beta\epsilon}{2}\right)
-{\rm tanh}\left(\frac{\beta(\epsilon+\omega)}{2}\right)\right)\nonumber\\
&\times& {\rm Im}(-i)\frac{1}{\Delta}\{ \frac{1}{u(\epsilon+\omega)-u(\epsilon)^*}
\frac{2}{\pi}
[\frac{u(\epsilon+\omega)}{\sqrt{1-u(\epsilon+\omega)^2}}
K\left(\frac{1}{\sqrt{1-u(\epsilon+\omega)^2}}\right)\nonumber\\
&-& \frac{u(\epsilon)^*}{\sqrt{1-u(\epsilon)^{*2}}}
K\left(\frac{1}{\sqrt{1-u(\epsilon)^{*2}}}\right)]\nonumber\\
&-& \frac{1}{u(\epsilon+\omega)-u(\epsilon)}\frac{2}{\pi}[
\frac{u(\epsilon+\omega)}{\sqrt{1-u(\epsilon+\omega)^2}}
K\left(\frac{1}{\sqrt{1-u(\epsilon+\omega)^2}}\right)\nonumber\\
&-& \frac{u(\epsilon)}{\sqrt{1-u(\epsilon)^2}}
K\left(\frac{1}{\sqrt{1-u(\epsilon)^2}}\right)]\}.
\end{eqnarray}
The imaginary part $\sigma_{2s}$ is also obtained from eq.(\ref{kfun}) as
\begin{equation}
\sigma_{2s}(\omega)= {\rm Im}\sigma_s(\omega)= 
\frac{1}{\omega}{\rm Re}K(q=0,\omega+i\delta),
\end{equation}
where Re$K(q=0,\omega+i\delta)$ is written as
\begin{eqnarray}
{\rm Re}K(q=0,\omega+i\delta)&&= \frac{e^2k_F^2}{2m^2}N(0)\int^{\infty}_{-\infty}
d\epsilon [(f(\epsilon)-f(\epsilon+\omega))\nonumber\\
&\times&{\rm Re}(\frac{1}{v(\epsilon+\omega)-v(\epsilon)^*}
\frac{2}{\pi}(K(\frac{\Delta}{v(\epsilon+\omega)})-
K(\frac{\Delta}{v(\epsilon)^*}))\nonumber\\
&&-\frac{1}{v(\epsilon+\omega)-v(\epsilon)}\frac{2}{\pi}
(K(\frac{\Delta}{v(\epsilon+\omega)})-K(\frac{\Delta}{v(\epsilon)})))\nonumber\\
&-&2f(\epsilon+\omega){\rm Re}{\frac{1}{v(\epsilon+\omega)-v(\epsilon)}
\frac{2}{\pi}(K(\frac{\Delta}{v(\epsilon+\omega)})-K(\frac{\Delta}{v(\epsilon)}))}]
\nonumber\\
&=&-\frac{e^2k_F^2}{2m^2}N(0)\int^{\infty}_{-\infty}d\epsilon[
\frac{1}{2}({\rm tanh}(\frac{\beta\epsilon}{2})-
{\rm tanh}(\frac{\beta(\epsilon+\omega)}{2}))\nonumber\\
&\times&{\rm Im}(\frac{1}{\Delta}\frac{1}{u(\epsilon+\omega)-u(\epsilon)^*}
\frac{2}{\pi}(\frac{u(\epsilon+\omega)}{\sqrt{1-u(\epsilon+\omega)^2}}
K(\frac{1}{\sqrt{1-u(\epsilon+\omega)^2}})\nonumber\\
&&-\frac{u(\epsilon)^*}{\sqrt{1-u(\epsilon)^{*2}}}K(\frac{1}{\sqrt{1-u(\epsilon)^{*2}}}))\nonumber\\
&-&\frac{1}{\Delta}\frac{1}{u(\epsilon+\omega)-u(\epsilon)}\frac{2}{\pi}
(\frac{u(\epsilon+\omega)}{\sqrt{1-u(\epsilon+\omega)^2}}
K(\frac{1}{\sqrt{1-u(\epsilon+\omega)^2}})\nonumber\\
&&-\frac{u(\epsilon)}{\sqrt{1-u(\epsilon)^2}}K(\frac{1}{\sqrt{1-u(\epsilon)^2}})))\nonumber\\
&-&2f(\epsilon+\omega){\rm Im}\frac{1}{\Delta}\frac{1}{u(\epsilon+\omega)-u(\epsilon)}
\frac{2}{\pi}
(\frac{u(\epsilon+\omega)}{\sqrt{1-u(\epsilon+\omega)^2}}
K(\frac{1}{\sqrt{1-u(\epsilon+\omega)^2}})\nonumber\\
&-&\frac{u(\epsilon)}{\sqrt{1-u(\epsilon)^2}}K(\frac{1}{\sqrt{1-u(\epsilon)^2}}))).
\end{eqnarray}

\subsection{Conductivity Sum Rule}
As shown in Section II.A, the sum rule holds for the conductivity:
\begin{equation}
\int^{\infty}_0 {\rm Re}\sigma(\omega)d\omega= \frac{\pi}{2}\frac{ne^2}{m},
\label{sigmasum}
\end{equation}
where $\sigma(\omega)$ is divided by the volume so that the quantity is
of the order of $O(1)$ and $n=N/V$ is the electron density.
In superconductors, there is a dramatic change in the optical conductivity
stemmed from opening of an excitation gap.  The change of the conductivity
is compensated by the formation of a zero frequency $\delta$ function peak
to preserve the conductivity sum rule.\cite{tin59}
From the general formula in eq.(\ref{sformula}), the real part of the
optical conductivity is written as
\begin{eqnarray}
{\rm Re}\sigma(\omega)&=& -{\rm Im}\frac{1}{\omega+i\delta}
(K({\bf q},\omega+i\delta)-K({\bf q},\omega\rightarrow 0))\nonumber\\
&=& -\frac{1}{\omega}{\rm Im}(K({\bf q},\omega+i\delta)-
K({\bf q},\omega\rightarrow 0))+\pi\delta(\omega){\rm Re}
(K({\bf q},\omega+i\delta)-K({\bf q},\omega\rightarrow 0))\nonumber\\
&=& -\frac{1}{\omega}{\rm Im}(K({\bf q},\omega+i\delta)-
K({\bf q},\omega\rightarrow 0))+\pi\delta(\omega)\omega{\rm Im}\sigma(\omega).
\end{eqnarray}
In superconductors, the first term indicates $\sigma_{1s}\equiv{\rm Re}\sigma$ 
in the finite
frequency region whose weight is removed by the opening of the gap, and the
second term represents the condensate peak which recovers the lost weight
for finite $\omega$.
The Drude weight is then given by 
$D=\pi\lim_{\omega\rightarrow 0}\omega{\rm Im}\sigma(\omega)$.  The sum rule
is expressed as
\begin{equation}
\int^{\infty}_0{\rm Re}\sigma(\omega)d\omega=-\int^{\infty}_0\frac{1}{\omega}
{\rm Im}(K({\bf q},\omega+i\delta)-K({\bf q},\omega\rightarrow 0))d\omega
+\frac{\pi}{2}\lim_{\omega\rightarrow 0}\omega{\rm Im}\sigma(\omega).
\label{sumsigma}
\end{equation}
If the total weight is conserved in the superconducting transition, we have
the relation from eq.(\ref{sumsigma}),
\begin{equation}
\int^{\infty}_0(\sigma_{1n}(\omega)-\sigma_{1s}(\omega))d\omega=
\frac{D}{2}=\frac{\pi}{2}\lim_{\omega\rightarrow 0}\omega\sigma_{2s}(\omega),
\label{sumrule}
\end{equation}
where $\sigma_{2s}$ denotes the imaginary part of $\sigma$ in the superconducting 
state.
In superconductors, the weight of the condensate peak can be estimated from the
measurements of the London penetration depth,
\begin{equation}
\lim_{\omega\rightarrow 0}\omega\sigma_{2s}(\omega)= 
\frac{c^2}{4\pi\lambda_L^2}=\frac{n_se^2}{m},
\end{equation} 
where the right-hand side is written as
$1/(\lambda_L^2\mu_0)$ in MKSA unit using the permeability of vacuum $\mu_0$.
As we will show in the next Section, the sum rule in eq.(\ref{sumrule}) actually holds
for the optical conductivity spectra $\sigma(\omega)$ of NbN$_{1-x}$C$_x$
obtained using the Reflectance-Transmittance method.\cite{shi02}

Recently, however, in cuprate superconductors, there is an experimental report
that the sum rule is violated for $c$-axis conductivity.\cite{bas99}
The weight estimated from the reflectance data is much
larger than can be accounted by the spectra integration in eq.(\ref{sumrule}).
This issue is not resolved at present whether the extra weight is coming
from outside the frequency range measured or coming from the lack of 
quasiparticle poles in the Green's functions.\cite{nor02}

At the last of this Section, we briefly discuss the ration between the
conductivity sum rule and the f-sum rule.  The dielectric function
$\epsilon(\omega)$ is related to the conductivity through the relation
\cite{noz99}
\begin{equation}
\epsilon(\omega)=1-\frac{4\pi}{i\omega}\sigma(\omega).
\end{equation}
The f-sum rule states that
\begin{equation}
\epsilon(\omega)\rightarrow 1-\frac{\omega^2_p}{\omega^2}~~~(\omega
\rightarrow\infty),
\end{equation}
where $\omega_p=\sqrt{4\pi ne^2/m}$ is the plasma frequency.  This indicates
\begin{equation}
\sigma(\omega)\rightarrow \frac{i}{\omega}\frac{ne^2}{m}~~~(\omega
\rightarrow\infty),
\end{equation}
and hence in the high-frequency limit ${\rm Im}\sigma(\omega)=(ne^2/m)/\omega$.
Since $\epsilon(\omega)$ is analytic in the upper half-plane, performing an
integration of $\omega(\epsilon(\omega)-1))$ in the upper-half plane, we obtain
\begin{equation}
\int^{\infty}_{-\infty}d\omega i4\pi\sigma(\omega)= i\pi\omega_p^2.
\end{equation}
This equation implies the sum rule in eq.(\ref{sigmasum}).

\section{Reflectance-Transmittance Method}
\subsection{Method of Analysis}
The conventional FIR spectroscopy based on
a Kramers-Kronig (K-K) transformation, however,
is rather unfavorable for studying electronic properties in small energy
region.\cite{tim89,gao93}
Recently, a new method to examine far-infrared (FIR) spectroscopy has been
developed without devoting to evaluating the Kramers-Kronig transformation.\cite{kog98}
In this method the optical conductivity is estimated from the data of 
reflectance spectra $R(\omega)$ and transmittance spectra $T(\omega)$ by
substituting them into a set of coupled equations.  The new method is free
from the conventional difficulties in far-infrared region since we do not
need the aid of Kramers-Kronig transformation.\cite{shi01}
This method is referred to as the R-T method since both $R(\omega)$ and $T(\omega)$
are necessary for analyses.
The basic concept of the R-T method was first applied to the study of $\sigma(\omega)$
of superconducting NbN thin films deposited on MgO and Si substrate for $\omega$
below $\sim 150$cm$^{-1}$ at $T=5-20$K.\cite{kog98,kaw02} 
In this section we briefly describe the concept of the R-T method and its advantages
over the conventional method based on K-K analysis using bulk samples.

The reflectance spectrum $R_1(\omega)$ and transmittance spectrum $T_1(\omega)$ of
a single-layered system (such as MgO substrate) are given by
\begin{equation}
R_1(\omega)= |r_1(\omega)|^2,
\label{r1}
\end{equation}
\begin{equation}
T_1(\omega)= |t_1(\omega)|^2,
\end{equation}
when the incident radiation is introduced in the direction normal to the layer
surface.  Here $r_1(\omega)$ and $t_1(\omega)$ are complex reflectance and
transmittance, respectively.  If the multiple internal reflections within the
layer are exactly taken into account, $r_1(\omega)$ and $t_1(\omega)$ are given by
\begin{equation}
r_1(\omega)=-\frac{(1-N_1)(N_1+1){\rm e}^{i(\omega/c)N_1d_1}+(1+N_1)(N_1-1)
{\rm e}^{-i(\omega/c)N_1d_1}}{(1-N_1)(N_1-1){\rm e}^{i(\omega/c)N_1d_1}+
(1+N_1)(N_1+1){\rm e}^{-i(\omega/c)N_1d_1}},
\end{equation} 
\begin{equation}
t_1(\omega)=-\frac{4N_1{\rm e}^{-i(\omega/c)d_1}}
{(1-N_1)(N_1-1){\rm e}^{i(\omega/c)N_1d_1}+
(1+N_1)(N_1+1){\rm e}^{-i(\omega/c)N_1d_1}},
\end{equation} 
where $c$ is the velocity of light in vacuum; $d_1$ is the thickness of the
layer; $N_1=n_1+ik_1$ is the complex refractive index.  
For a two-layered system (such as YBCO thin films deposited on MgO
substrate) placed in vacuum, the reflectance $R_2(\omega)$ and transmittance
$T_2(\omega)$ are given by
\begin{equation}
R_2(\omega)= |r_2(\omega)|^2,
\end{equation}
\begin{equation}
T_2(\omega)= |t_2(\omega)|^2,
\end{equation}
where we assume the same conditions for incident radiation.
$r_2(\omega)$ and $t_2(\omega)$ are given by
\begin{equation}
r_2(\omega)= \frac{A+B+C+D}{E+F+G+H},
\end{equation}
\begin{equation}
t_2(\omega)= \frac{J}{E+F+G+H},
\end{equation}
where
\begin{equation}
A= -(N_1-N_2)(N_2+1)(N_1+1){\rm e}^{i(\omega/c)(N_2d_2-N_1d_1)},
\end{equation}
\begin{equation}
B= -(N_1+N_2)(N_2-1)(N_1+1){\rm e}^{-i(\omega/c)(N_2d_2+N_1d_1)},
\end{equation}
\begin{equation}
C= (N_1+N_2)(N_2+1)(N_1-1){\rm e}^{i(\omega/c)(N_2d_2+N_1d_1)},
\end{equation}
\begin{equation}
D= (N_1-N_2)(N_2-1)(N_1-1){\rm e}^{-i(\omega/c)(N_2d_2-N_1d_1)},
\end{equation}
\begin{equation}
E= (N_1-N_2)(N_2-1)(N_1+1){\rm e}^{i(\omega/c)(N_2d_2-N_1d_1)},
\end{equation}
\begin{equation}
F= (N_1+N_2)(N_2+1)(N_1+1){\rm e}^{-i(\omega/c)(N_2d_2+N_1d_1)},
\end{equation}
\begin{equation}
G=-(N_1+N_2)(N_2-1)(N_1-1){\rm e}^{i(\omega/c)(N_2d_2+N_1d_1)},
\end{equation}
\begin{equation}
H=-(N_1-N_2)(N_2+1)(N_1-1){\rm e}^{-i(\omega/c)(N_2d_2-N_1d_1)},
\end{equation}
\begin{equation}
J= 8N_1N_2{\rm e}^{-i(\omega/c)(d_2+d_1)}.
\label{eqj}
\end{equation}
Here $d_2$ and $N_2=n_2+ik_2$ are the thickness and the complex refractive index of
thin films, respectively.\cite{kog98,kaw02}
The equations (\ref{r1})-(\ref{eqj}) are basic relations in R-T method.

The coupled equations are numerically solved to determine the values of complex
refractive indices $n_i$ and $k_i$ ($i=1,2$) as functions of $\omega$.
First, the coupled equations for $T_1(\omega)$ and $R_1(\omega)$ measured for
the MgO substrate are solved using the Newton method, then we obtain $n_1$ and
$k_1$ as a function of $\omega$.  Second, $T(\omega)$ and $R(\omega)$ of
YBCO/MgO are measured.  The measured values are substituted into $R_2$ and $T_2$
for a given value, as well as $n_1$ and $k_1$ for the MgO substrate obtained in
the first procedure.  Then $n_2$ and $k_2$ of YBCO are determined as a function of
$\omega$ by solving the coupled equations numerically.

Here, we briefly discuss both the advantages and disadvantages of the R-T
method in comparison with the conventional method based on the K-K analysis using 
bulk samples.  First, we can estimate precisely the optical constants of materials
such as the complex refractive index.  Second, the accuracy of $\sigma_1$ obtained
by the R-T method is expected to be better than that obtained using the conventional
method in the low frequency region.  We have three reasons for this expectation.
(i) In the conventional method we use only $R(\omega)$, while in the R-T method
both $R(\omega)$ and $T(\omega)$ are employed to obtain the complex refractive
index.
(ii) We must be careful about the K-K analysis in the conventional method.
For metallic materials, $R(\omega)$  takes high values in the small $\omega$
region and thus the experimental errors increase through the K-K transformation.
The errors in $R(\omega)$ propagate to the errors in $\sigma_1$.  The conventional
method is not so reliable in the study of $\sigma_1(\omega)$ in the small $\omega$
region for metallic materials.
(iii) The K-K analysis usually requires an extrapolation of $R(\omega)$ to the
low- and high-frequency regions beyond the measurable $\omega$ region.
Since there are no guiding principles for the extrapolation of $R(\omega)$,
there may be some uncertainty in the K-K transformation in the conventional method.
In particular, the ambiguities in the extrapolation are not negligible in the low 
frequency region.
The R-T method does not have this kind of difficulty.

A disadvantage of the R-T method lies in the fact that the range of $\omega$
for which the method is applicable is limited because the substrate must be
transparent to the incident radiation to measure $T(\omega)$.
For MgO as the substrate, the transparent range is about $0-300$ cm$^{-1}$
in far-infrared region even at $T$ below 10 K.\cite{jas66,tyang90}
In addition, the high-frequency limit of the transparent range decreases steeply
as $T$ increases for MgO; this limit is approximately 100 cm$^{-1}$ at room
temperature.  Since the conventional method based on the K-K transformation is
expected to work very well for $\omega > 300$cm$^{-1}$, the region of $\omega$
within the scope of the R-T method is $\omega < 300$cm$^{-1}$.

\begin{figure}[bp]
\includegraphics[width=12cm]{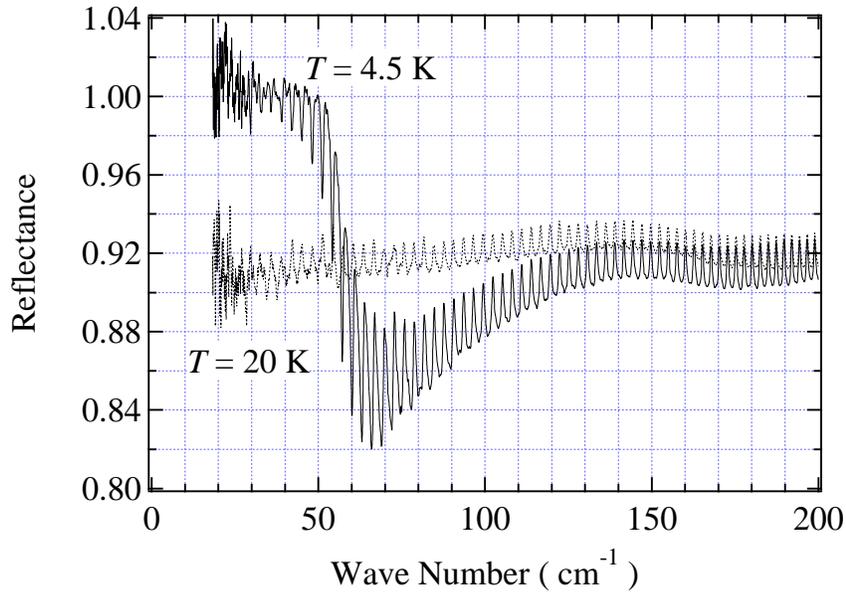}
\caption{
Reflectance spectra $R(\omega)$ of NbN$_{1-x}$C$_x$ thin films deposited on MgO 
substrates at $T=4.3$ K (solid line) and $T=20$ K (dotted line).
}
\label{nbnref}
\end{figure}

\begin{figure}[bp]
\includegraphics[width=12.5cm]{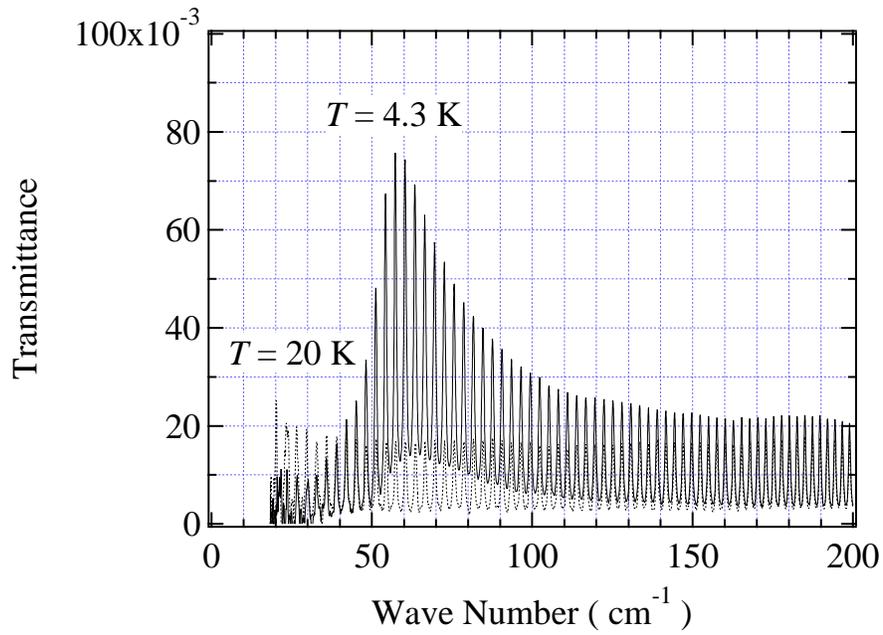}
\caption{
Transmittance spectra $T(\omega)$ of NbN$_{1-x}$C$_x$ thin films deposited on MgO 
substrates at $T=4.3$ K (solid line) and 20 K (dotted line).
}
\label{nbntra}
\end{figure}

\begin{figure}[bp]
\includegraphics[width=12cm]{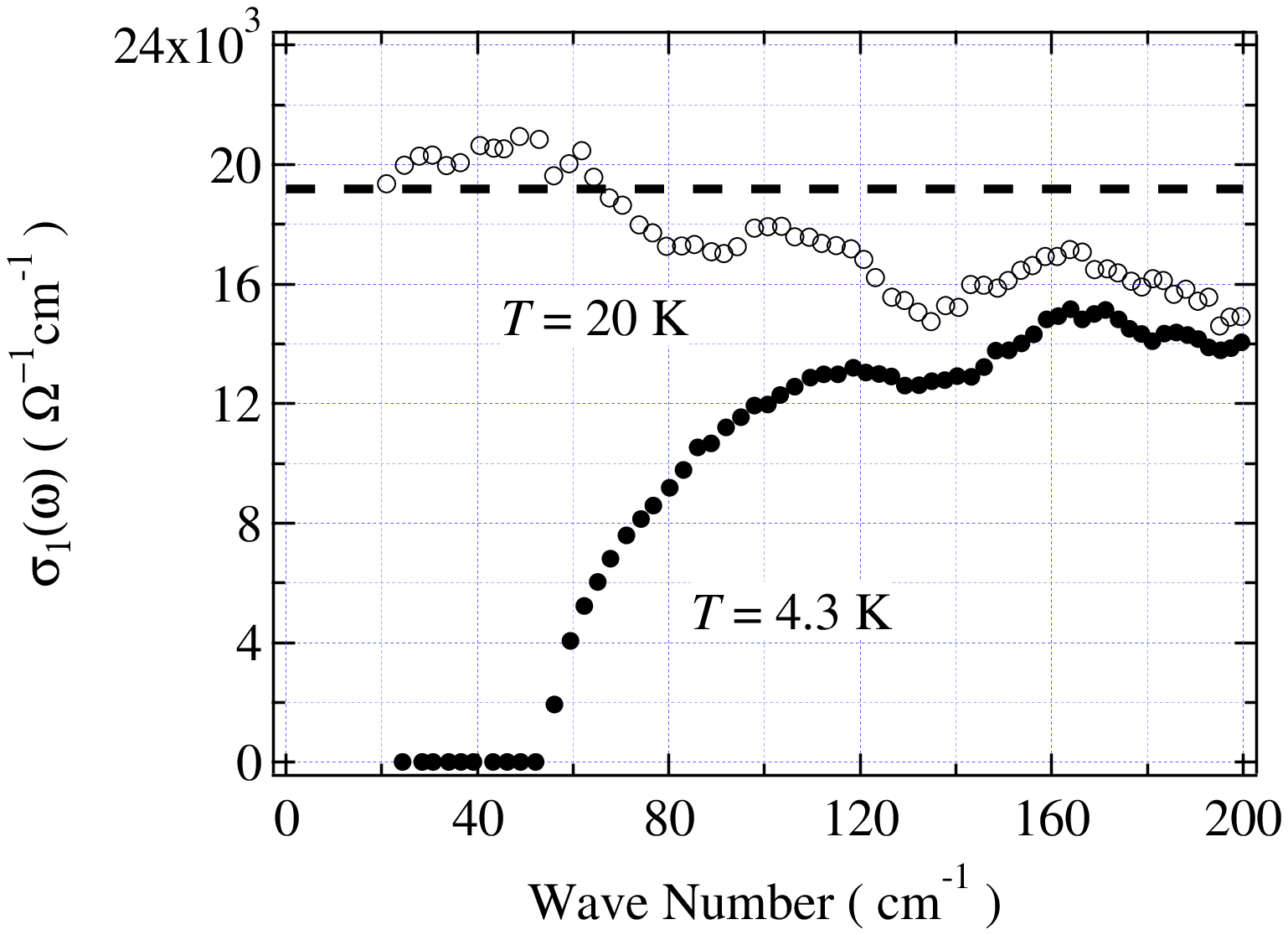}
\caption{
$\sigma_1(\omega)$ for NbN$_{1-x}$C$_x$ calculated by the R-T method at
$T=4.3$ K (solid circles) and $T=20$ K (open circles).
The dashed line shows the results of calculations using the Drude formula
at $T=20$ K.
}
\label{nbns1}
\end{figure}

\begin{figure}[bp]
\includegraphics[width=12cm]{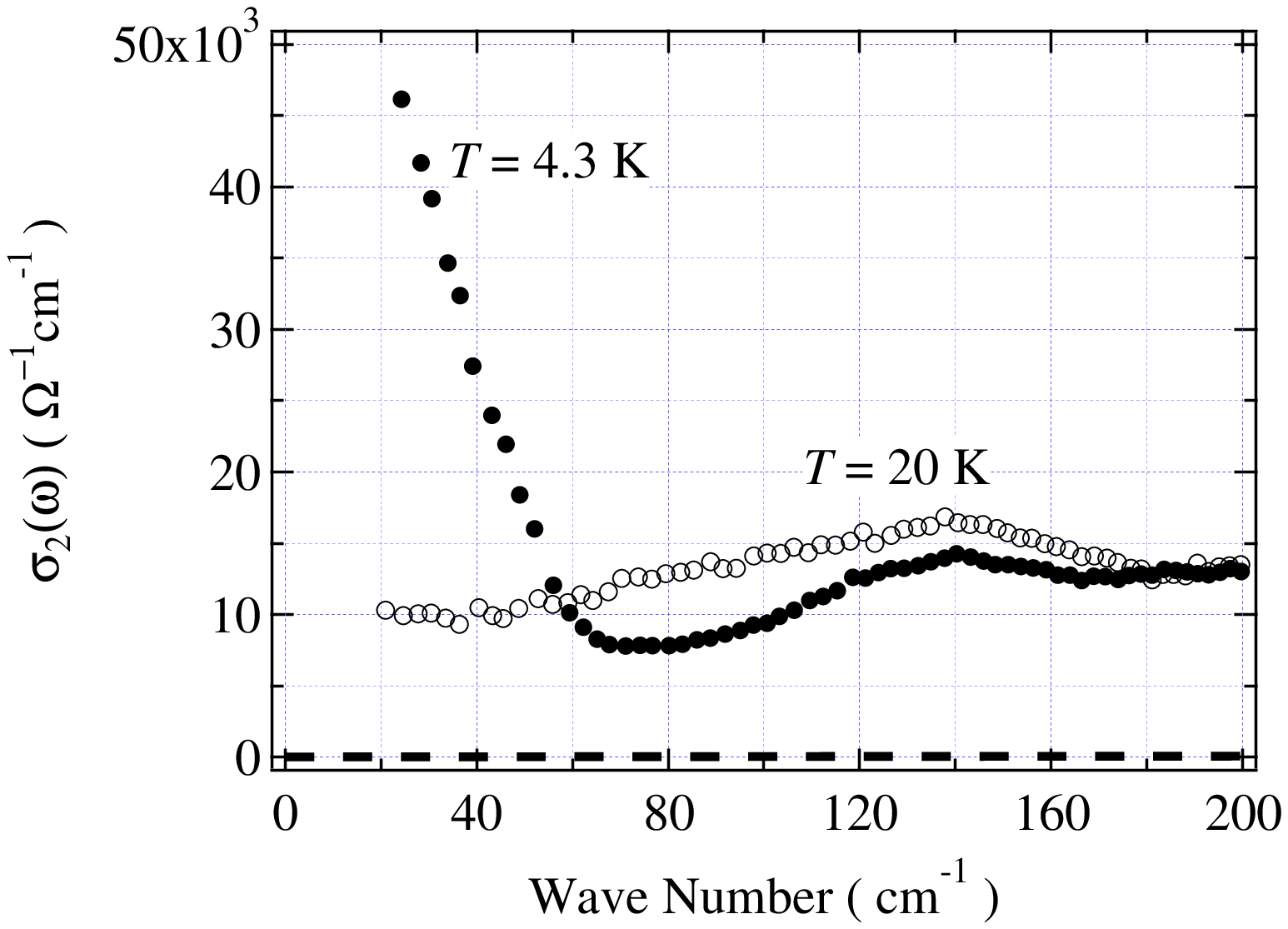}
\caption{
$\sigma_2(\omega)$ for NbN$_{1-x}$C$_x$ calculated by the R-T method at
$T=4.3$ K (solid circles) and $T=20$ K (open circles).
The dashed line shows the results of calculations using the Drude formula
at $T=20$ K.
}
\label{nbns2}
\end{figure}

\begin{figure}[bp]
\includegraphics[width=12cm]{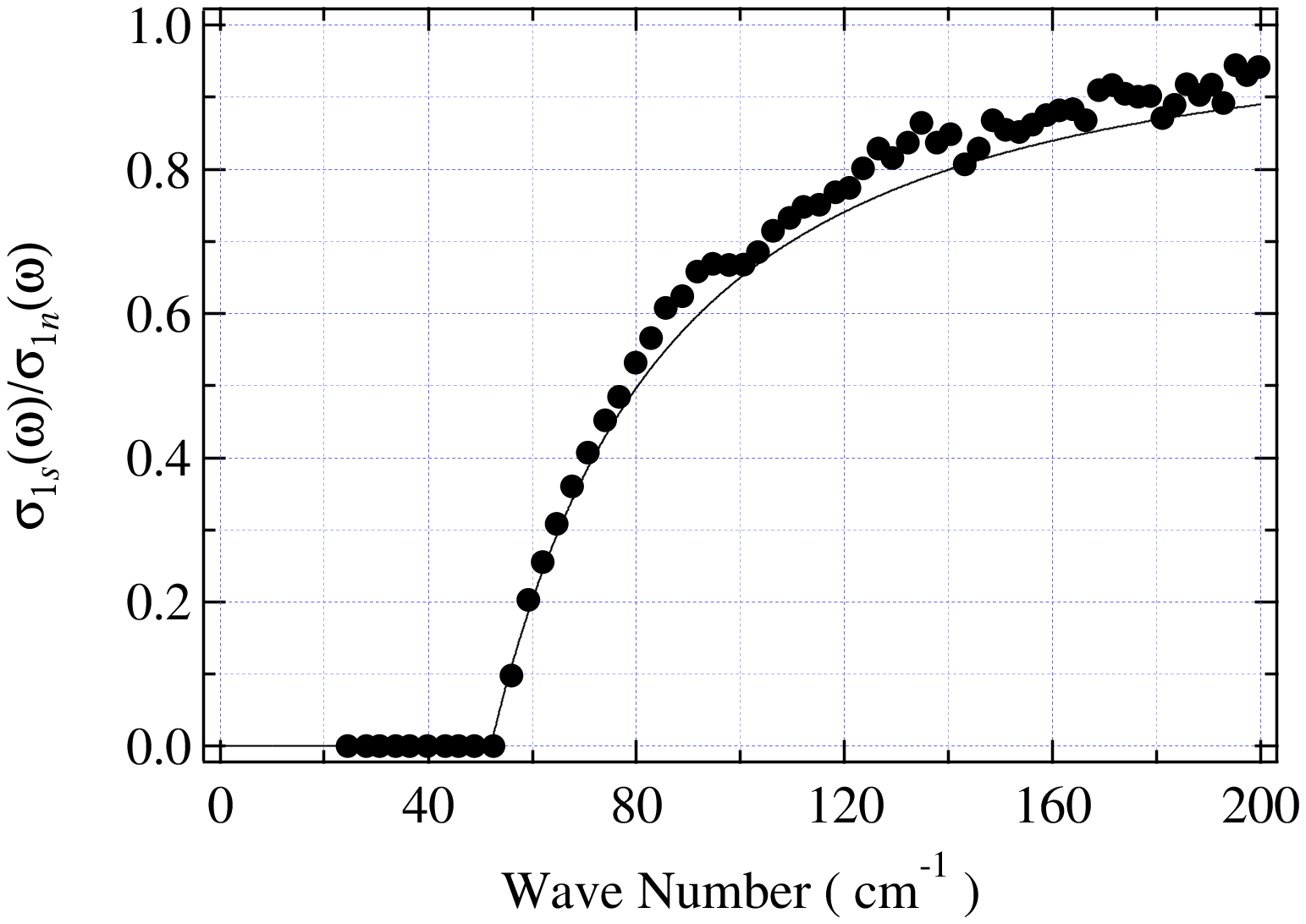}
\caption{
The relative conductivity ratio $\sigma_{1s}(\omega)/\sigma_{1n}(\omega)$ where
$\sigma_1(\omega)$ at $T=4.3$ K is divided by $\sigma_1(\omega)$ at $T=20$ K.
The solid line shows the results obtaned from the Mattis-Bardeen formula.
}
\label{nbns1s}
\end{figure}

\begin{figure}[bp]
\includegraphics[width=12cm]{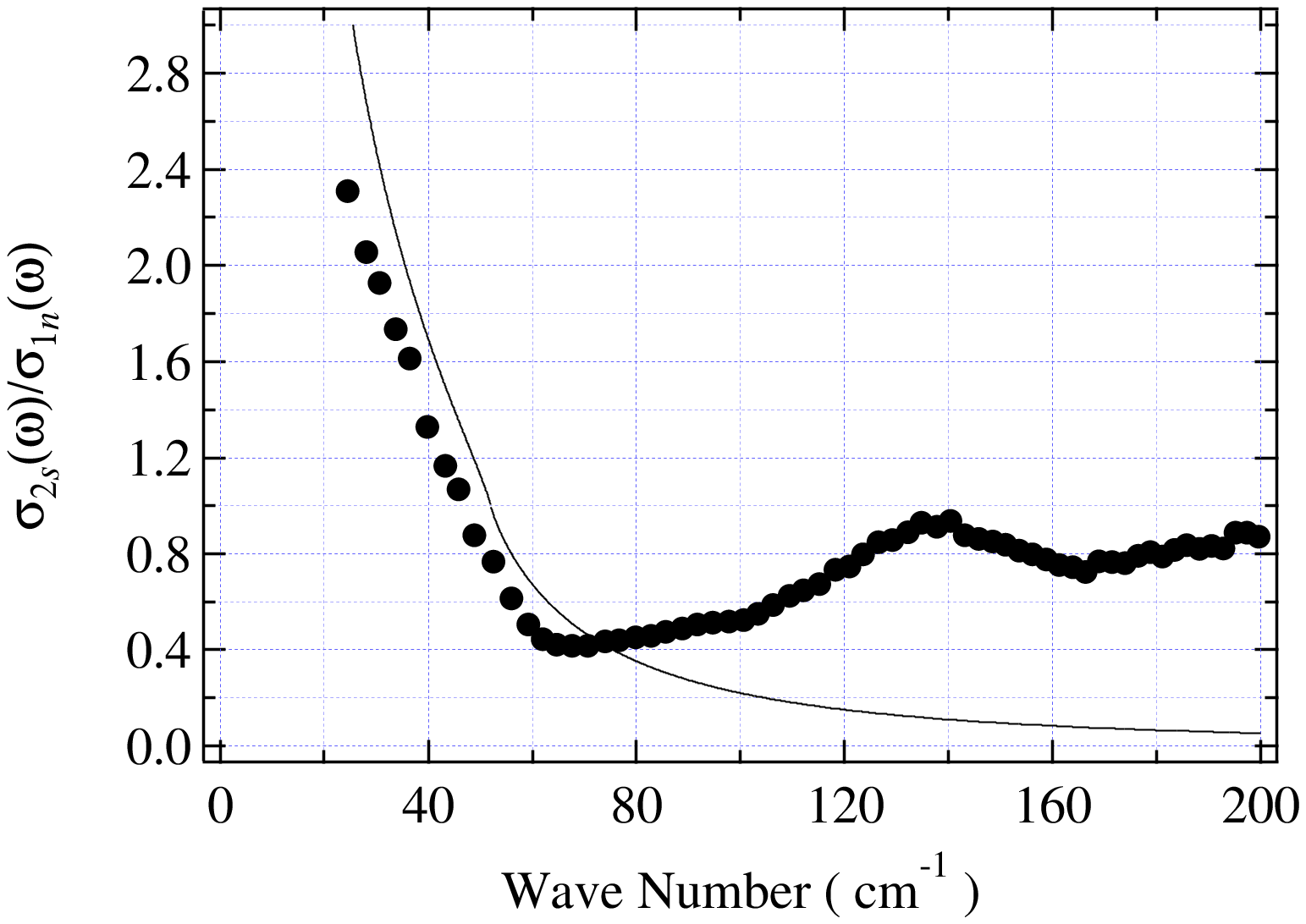}
\caption{
The relative conductivity ratio $\sigma_{2s}(\omega)/\sigma_{2n}(\omega)$ where
$\sigma_2(\omega)$ at $T=4.3$ K is divided by $\sigma_2(\omega)$ at $T=20$ K.
The solid line shows the results obtaned from the Mattis-Bardeen formula.
}
\label{nbns2s}
\end{figure}

\subsection{Experimental Results}
In order to examine the feasibility of the R-T method, we report $R(\omega)$ and
$T(\omega)$ of NbN$_{1-x}$C$_x$ thin films deposited on MgO substrates
in the far-infrared region.
The substrates were 0.5 mm thick.  The thickness of the NbN$_{1-x}$C$_x$ layers
was about 40 nm.  The films were epitaxial and $T_c$ was estimated to be 17.5 K
by electrical measurements.  The value of $x$ was estimated to be less than 0.3.

The electrical resistivity measured by the four-probe method was
$\rho=5.2\times 10^{-5}$ $\Omega$cm at $T=20$ K.  The carrier density in the
normal state $n$ was measured by the van der Pauw method:
$n=1.29\times 10^{23}$ cm$^{-3}$ which is slightly lower than the value of
$n=2.39\times 10^{23}$ cm$^{-3}$ reported previously for NbN\cite{math72}.
The carrier scattering rate $\Gamma$ at $T=20$ K was estimated as
$\Gamma=\rho e^2n/m^*=1.9\times 10^{15}$ s$^{-1}$ = 6.3$\times 10^4$ cm$^{-1}$,
where the effective mass $m^*$ is assumed to be equal to the electron rest mass.
The superconducting gap has been estimated as $2\Delta\approx 50$ cm$^{-1}$.
Thus NbN$_{1-x}$C$_x$ is suggested to be a typical dirty-limit BCS superconductor.

$R(\omega)$ and $T(\omega)$ obtained at $T=4.3$ K and 20 K are shown in 
Figs. \ref{nbnref} and \ref{nbntra}. 
The interference fringes due to multiple internal reflections within the
MgO substrate are clearly visible in both figures because the MgO substrate is
highly transparent in this $\omega$ region at $T=4.3$ K and 30 K, and because
the NbN$_{1-x}$C$_x$ film is thin enough to transmit far-infrared radiation.
$R(\omega)$ at $T=4.3$ K exhibits an obvious reflectance edge at $\omega\sim 65$
cm$^{-1}$ and $R(\omega)\sim 1$ for $\omega$ less than the reflectance edge
frequency, which is a special characteristic for superconductors.
$T(\omega)$ at $T=4.3$ K exhibits a maximum at $\omega\sim 60$ cm$^{-1}$, which
is related to the evolution of the reflectance edge in $R(\omega)$.

We show $\sigma_1(\omega)$ and $\sigma_2(\omega)$ spectra Figs. \ref{nbns1} and 
\ref{nbns2} for 
NbN$_{1-x}$C$_x$
calculated by the R-T method using the experimental results shown in Figs. 
\ref{nbnref} and \ref{nbntra}.
The value of the dc conductivity $\sigma_1(0)$ at $T=20$ K is estimated to be
$\sigma_1(0)\sim 2.0\times 10^4$ $\Omega^{-1}$cm$^{-1}$ from Fig. \ref{nbns1}; 
this value
agrees well with the value of $1/\rho=1.9\times 10^4$ $\Omega^{-1}$cm$^{-1}$
estimated from the electrical measurements.
 
We evaluated the relative conductivity ratio 
$\sigma_{1s}(\omega)/\sigma_{1n}(\omega)$ and
$\sigma_{2s}(\omega)/\sigma_{1n}(\omega)$ from the results in Figs. \ref{nbns1} and 
\ref{nbns2},
where $\sigma_{1n}$ is $\sigma_1$ at $T=20$ K and $\sigma_{1s}$ and 
$\sigma_{2s}$ are at $T=4.3$ K, respectively.  The real and imaginary parts
of the relative conductivity
ratios are shown in Figs. \ref{nbns1s} and \ref{nbns2s}, respectively.  Here 
theoretical curves
obtained using the Mattis-Bardeen theory are also shown by solid lines, where
we set $2\Delta=52$ cm$^{-1}$ in accord with the value reported by the junction
method\cite{koh99}.  The ratio of $2\Delta$ to $T_c$ is given by
$2\Delta/k_BT_c\sim 4.3$, suggesting the strong-coupling superconductivity
in NbN$_{1-x}$C$_x$. 

The experimental results for $\sigma_{1s}/\sigma_{1n}$ in Fig. \ref{nbns1s} exhibit an
excellent agreement with the Mattis-Bardeen theory, and 
$\sigma_{2s}/\sigma_{1n}$ also shows a good agreement for $\omega$ less than
$\sim 60$ cm$^{-1}$ as shown in Fig. \ref{nbns2s}.  The agreement is, however, poor for
$\omega$ larger than $\sim 70$ cm$^{-1}$ in Fig. \ref{nbns2s}; this anomalous behavior
may be due to impurities\cite{zim91}.

Now we investigate the conductivity sum rule in eq.(\ref{sumrule}).
From the integration of spectra $\sigma_{1n}$ and $\sigma_{1s}$, 
$\lambda_L$ was estimated to be $\sim 193$ nm, while the $\sigma_{2s}$ spectra
in the superconducting state gives $\lambda_L\sim 200$ nm.  These values
show an excellent agreement, and also agrees well with the value reported
previously.\cite{koh99}  This indicates that the sum rule holds for
NbN$_{1-x}$C$_x$.

\begin{figure}[bp]
\includegraphics[width=12cm]{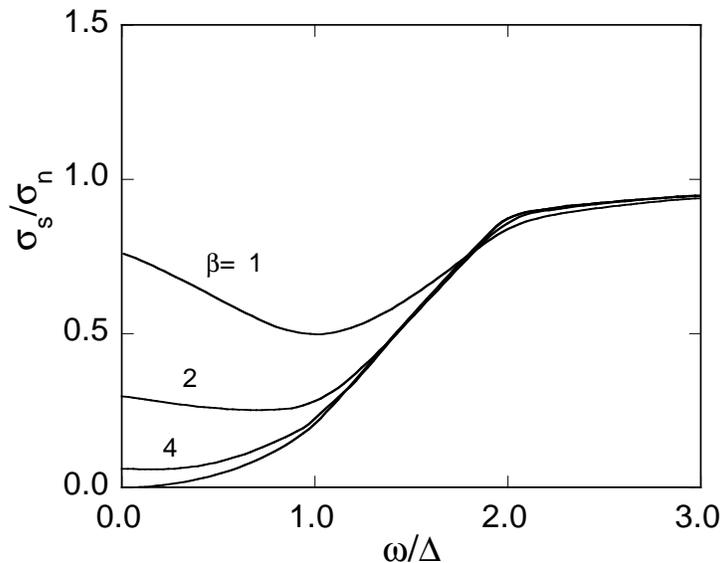}
\caption{
Real part of the optical conductivity as a function of $\omega$ for several values of
temperature.  From the top $T/\Delta=1/\beta=1,1/2,1/4$ and 0.
}
\label{dwave}
\end{figure}

\begin{figure}[bp]
\includegraphics[width=11cm]{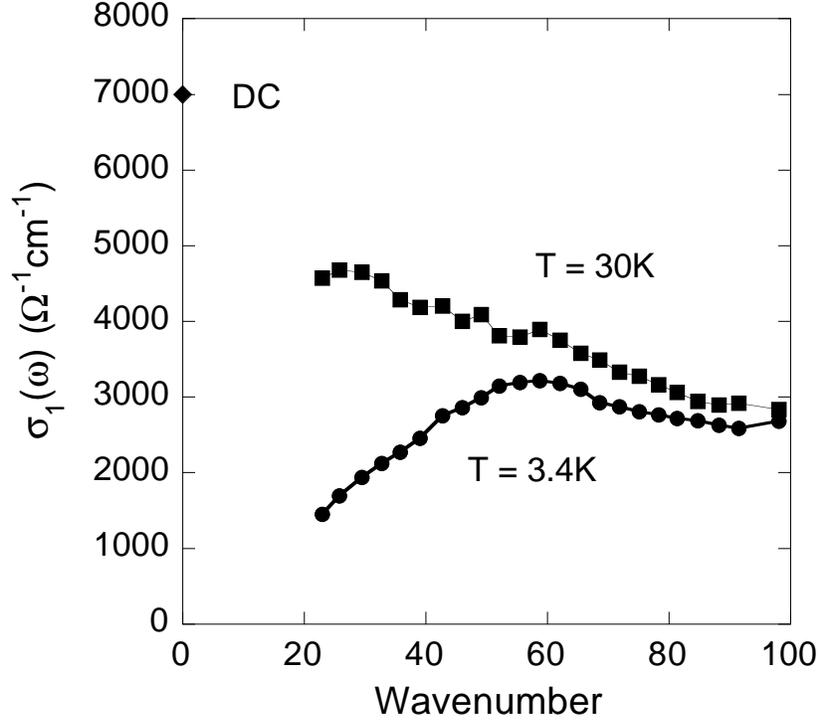}
\caption{
Optical conductivity calculated from the R-T method at $T=4.3$K (circles)
and 30K (squares).  The diamond indicates the DC value at $T=30$ K.
}
\label{sigma1Ex}
\end{figure}

\begin{figure}[bp]
\includegraphics[width=11cm]{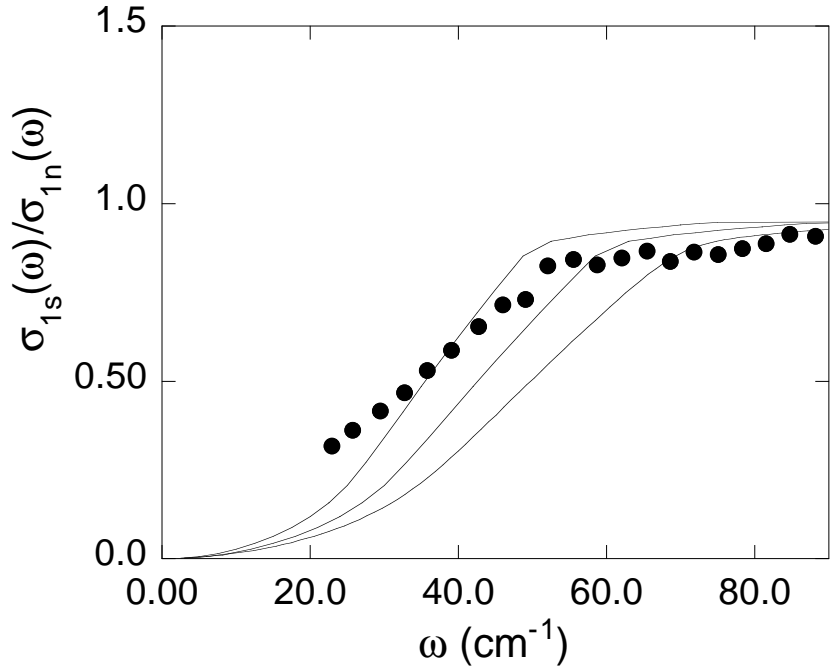}
\caption{
Optical conductivity (circles) by the R-T method and
theoretical predictions at $T=0$ (solid curves).  Form the left
$2\Delta=50$cm$^{-1}$, 60$cm^{-1}$ and 70cm$^{-1}$
}
\label{sigma2}
\end{figure}

\section{Electron-doped high-$T_c$ superconductor: London limit}
Oxide high-$T_c$ superconductors have been investigated intensively over the
last decade.  The $d$-wave superconductivity is well 
established for hole-doped superconductors.
However, there is a class of high-$T_c$
superconductors doped with electrons,\cite{tok89,tak89} for which both 
$s$-wave\cite{kas98} and $d$-wave pairing\cite{tsu00,kok00,pro00} have been 
reported.
Nd$_{2-x}$Ce$_x$CuO$_4$ is a typical example
of electron-doped materials and the symmetry of Cooper pairs has been 
controversial.
It is important to examine the symmetry of Cooper pairs in the study of
high-$T_c$ superconductors.

Since the superconducting gap $\Delta$ in 
Nd$_{2-x}$Ce$_x$CuO$_4$
is very small, there have been no reports on the study of the nature of the
superconducting gap of Nd$_{2-x}$Ce$_x$CuO$_4$ through such techniques,
although there have been a number of reports on the FIR spectroscopy of
Nd$_{2-x}$Ce$_x$CuO$_4$.\cite{hom97,ono99}
 
The purpose of this paper is to investigate FIR optical properties of 
Nd$_{2-x}$Ce$_x$CuO$_4$ obtained by the R-T method from a viewpoint of 
unconventional
superconductors.  We will show that the available data for the optical
conductivity and transmittance are well explained by
$d$-wave pairing model in the clean limit.  The value of superconducting gap
is estimated as $2\Delta\sim 60-70$ cm$^{-1}$, which is consistent with the
available value estimated by scanning tunneling spectroscopy.\cite{kas98}

\begin{figure}[bp]
\includegraphics[width=11.5cm]{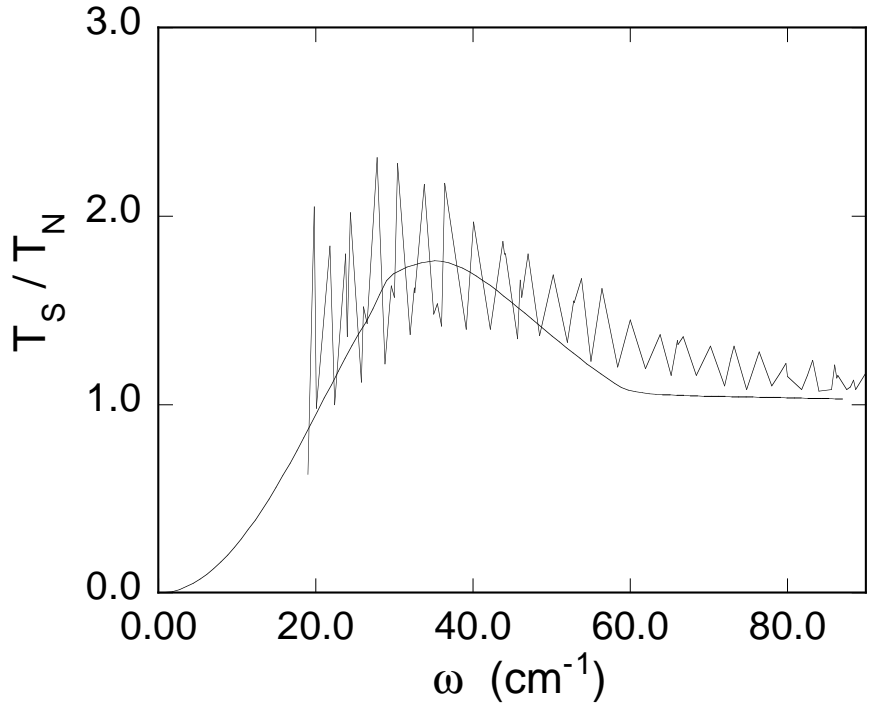}
\caption{
Observed Transmittance and the 
theoretical curves at $T=0$ (solid curve) for
$2\Delta=60$cm$^{-1}$.
}
\label{tstn}
\end{figure}

The frequency-dependent conductivity $\sigma(\omega)$ was calculated by
Mattis and Bardeen,\cite{mat58} Abrikosov {\em et al.}\cite{abr59} and 
Skalski {\em et al.}\cite{ska64} for
isotropic superconductors.  The original Mattis-Bardeen theory was carried
through for a conventional type-I $s$-wave superconductor, where the coherence
length $\xi$ and magnetic penetration depth $\lambda$ satisfy $\xi\gg\lambda$.
The opposite limit $\xi\ll\lambda$ (London limit) was also examined for 
$s$-wave
pairing by field theoretical treatments.\cite{ska64}
For the high-$T_c$ compounds of type-II superconductor with small coherence 
length, 
the formula in the London limit is appropriate for optical conductivity
measurements.  Recently the conductivity $\sigma(\omega)$ of an
unconventional superconductor has been examined theoretically in the London 
limit.\cite{hir89,hir92,hir94,gra95}
We use the current response function shown in Section II:
\begin{equation}
K_{\mu\nu}({\bf q},i\omega_m)= \frac{e^2k_F^2}{m^2}\sum_k \hat{k}_{\mu}\hat{k}_{\nu}
\frac{1}{\beta}\sum_n {\rm Tr}[G({\bf k}_+,i\epsilon_n+i\omega_m)
G({\bf k}_-,i\epsilon_n)],
\end{equation}
where ${\bf k}_{\pm}={\bf k}\pm {\bf q}/2$ and $\epsilon_n=(2n+1)\pi/\beta$.
The single-particle matrix Green's function is
\begin{equation}
G({\bf k},i\epsilon_n)= \frac{i(\epsilon_n-\Sigma(\epsilon_n))\tau^0+\xi_k\tau^3+\Delta_k\tau^1}{(\epsilon_n-\Sigma(\epsilon_n))^2+\xi_k^2+\Delta_k^2},
\end{equation}
where $\Delta_k$ is the anisotropic order parameter and $\Sigma(\epsilon_n)$ 
is the
self-energy due to impurity scattering. 
$\tau^i$ ($i=0,1,\cdots$) denote Pauli matrices.
Since we consider the case where $\xi\ll\lambda$ holds, the real part of
optical conductivity is well approximated by the formula in the London limit:
\begin{equation}
\sigma_{1s,\mu\nu}(\omega)= -\frac{1}{\omega}\lim_{q\rightarrow 0}{\rm Im}
K_{\mu\nu}({\bf q},\omega).
\end{equation}
Our focus is the collision less limit of the normalized conductivity to
compare it with the data for Nd$_{2-x}$Ce$_x$CuO$_4$ since $\xi\ll\ell$ holds
for the mean-free path $\ell$.
For anisotropic superconducting order parameter $\Delta_k$ such that 
the average over the Fermi surface vanishes $\langle\Delta_k\rangle=0$,
the expression for $\sigma_{1s}\equiv\sigma_{1s,xx}(\omega)$ in the collision less 
limit on the plane
is simply given by\cite{hir92}
\begin{equation}
\frac{\sigma_{1s}(\omega)}{\sigma_{1n}(\omega)}= \frac{1}{2\omega}
\int_{-\infty}^{\infty}dx\langle{\rm Re}
\frac{|x|}{\sqrt{x^2-\Delta_k^2}}\rangle
\langle{\rm Re}\frac{|x-\omega|}{\sqrt{(x-\omega)^2-\Delta_k^2}}\rangle
[{\rm tanh}(\frac{\beta x}{2})-{\rm tanh}(\frac{\beta(x-\omega)}{2})],
\end{equation}
which is an angle-dependent generalization of the Mattis-Bardeen formula.
For the $d$-wave symmetry, the average over the Fermi surface denoted by the
angular brackets is defined as
\begin{equation}
\langle{\rm Re}\frac{x}{\sqrt{x^2-\Delta_k^2}}\rangle=
{\rm Re} \int \frac{d\phi}{2\pi}\frac{x}{\sqrt{x^2-(\Delta{\rm cos}(2\phi))^2}},
\end{equation}
where the order parameter is factorized as $\Delta_k=\Delta{\rm cos}(2\phi)$.
In Fig.\ref{dwave} we show the behaviors of 
$\sigma_{1s}/\sigma_{1n}$
as a function of $\omega$ for several values of temperature $T$.
The infrared behaviors reflect the lines of nodes on the Fermi surface.

FIR reflection $R(\omega)$ and transmission $T(\omega)$ measurements were 
performed for
Nd$_{2-x}$Ce$_x$CuO$_4$ ($x=0.15$) thin films deposited by laser ablation 
onto (001)  MgO substrates.  The thickness of Nd$_{2-x}$Ce$_x$CuO$_4$ thin film
was about 40 nm.  
$T_c$ was estimated to be $\sim$20K.  The electric field
of the FIR radiation was predominantly parallel to the $a$-$b$ plane.
The conductivity spectra were evaluated by the R-T method from the data
for $R(\omega)$ and $T(\omega)$ at $T=4.3$ and 30K.\cite{shi01}

\begin{figure}[bp]
\includegraphics[width=10cm]{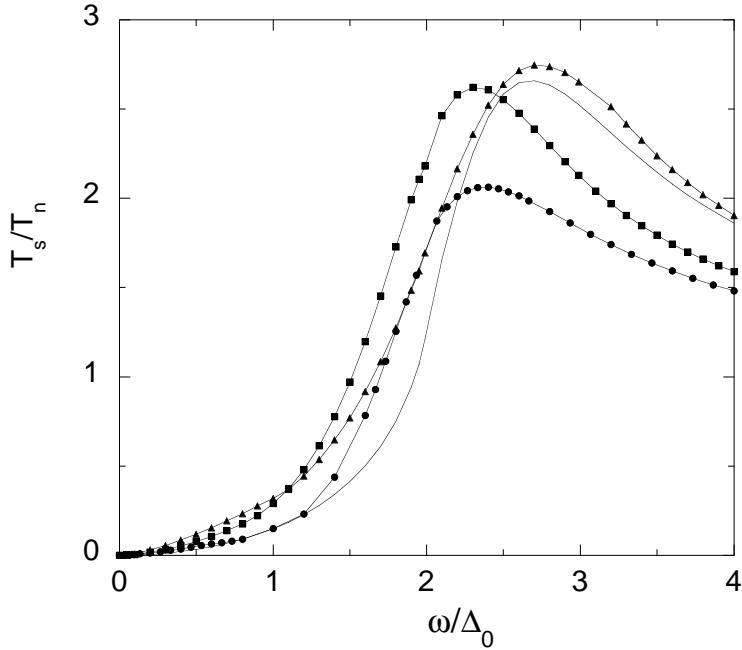}
\caption{
Transmission $T_S/T_N$ for anisotropic $s$-wave models.
The solid curve without marks is the Mattis-Bardeen result.
Squares, triangles and circles are for the prolate ($a=0.5$), ab-plane anisotropic
($c=0.5$), and oblate ($a=-0.5$)
gaps, respectively.  The oblate form shows a small increase compared to other types.
For the oblate gap, $\Delta_0=\Delta_{max}$, and for other types, $\Delta_0=\Delta$.
}
\label{tstn1}
\end{figure}
 
\begin{figure}[bp]
\includegraphics[width=12cm]{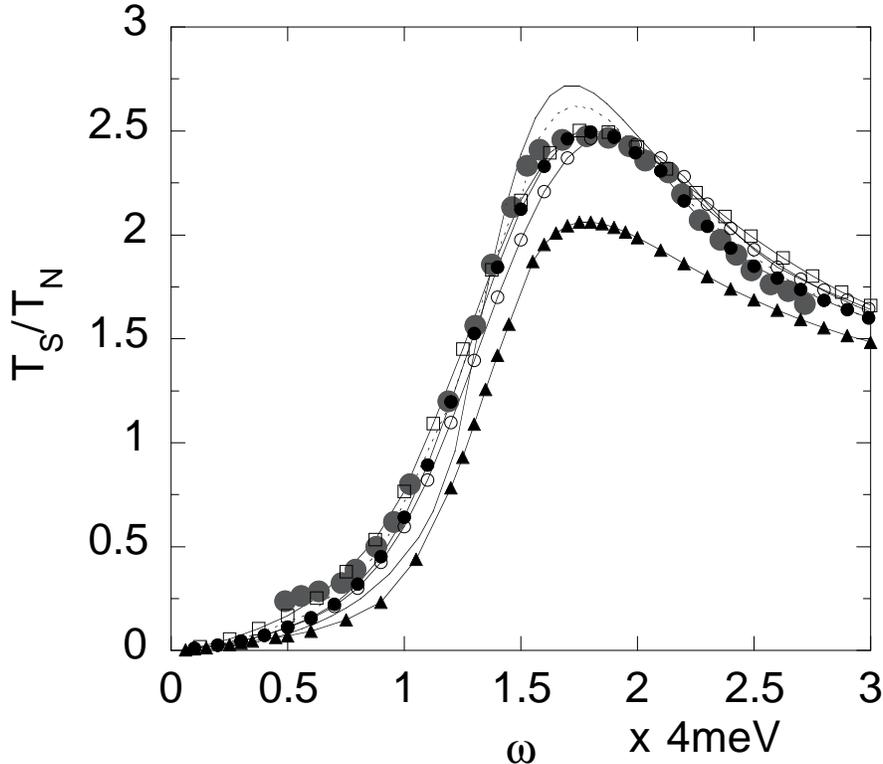}
\caption{
Transmission $T_S/T_N$ for the two-band model.
The data points (large solid circles) are taken from ref.\cite{kai02} at $T=6$K.
The solid line without marks shows the Mattis-Bardeen result with $2\Delta=5$meV.
The dotted line is for the single gap of prolate type with $2\Delta_{max}=9$meV and
$a=0.5$.  Triangles are for the single-gap model of oblate form with
$2\Delta_{max}=9$meV and $a=-0.33$.
Others are for the two-band gap model where the ab-plane anisotropy for
the $\sigma$-band
and the prolate form for the $\pi$-band are assumed.  The parameters are the following.
Solid circles: $2\Delta_{max}=8.5$meV and $c=0.33$ ($\sigma$-band: weight 0.45);
$2\Delta_{max}=6$meV and $a=0.33$ ($\pi$-band: weight 0.55).
Open circles: $2\Delta_{max}=9$meV and $c=0.33$ ($\sigma$-band: weight 0.45);
$2\Delta_{max}=6$meV and $a=0.33$ ($\pi$-band: weight 0.55).
Squares: $2\Delta_{max}=10$meV and $c=0.5$ ($\sigma$-band: weight 0.4);
$2\Delta_{max}=7.5$meV and $a=0.5$ ($\pi$-band: weight 0.6).
}
\label{tstn2}
\end{figure}

\begin{figure}[bp]
\includegraphics[width=10cm]{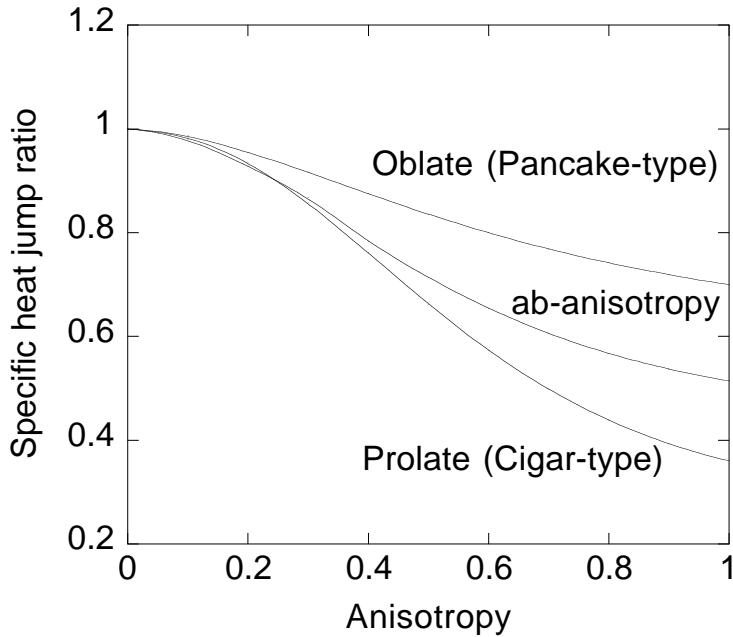}
\caption{
Specific-heat-jump ratio to the BCS value.  From the top the ratios for the oblate,
$ab$-anisotropy, and prolate gaps are shown, respectively.
}
\label{spec}
\end{figure}

\begin{figure}[bp]
\includegraphics[width=10cm]{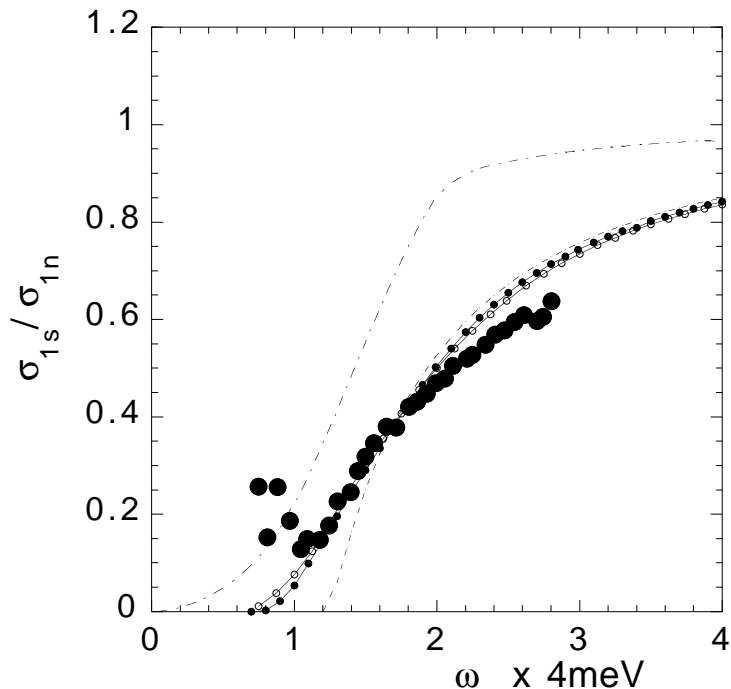}
\caption{
Real part of the optical conductivity for the two-band anisotropic model.
The data points (large solid circles) are taken from ref.\cite{kai02}
The parameters for solid circles are
$2\Delta_{max}=8.5$meV and $c=0.33$ ($\sigma$-band: weight 0.45);
$2\Delta_{max}=6$meV and $a=0.33$ ($\pi$-band: weight 0.55).
The parameters for open circles are
$2\Delta_{max}=10$meV and $c=0.5$ ($\sigma$-band: weight 0.4);
$2\Delta_{max}=7.5$meV and $a=0.5$ ($\pi$-band: weight 0.6).
The dashed line indicates the results obtained using the Mattis-Bardeen formula
with $2\Delta=5$meV.
The dash-dotted line denotes the conductivity for the $d$-wave gap.\cite{yana01}
}
\label{s1}
\end{figure}

\begin{figure}[bp]
\includegraphics[width=10cm]{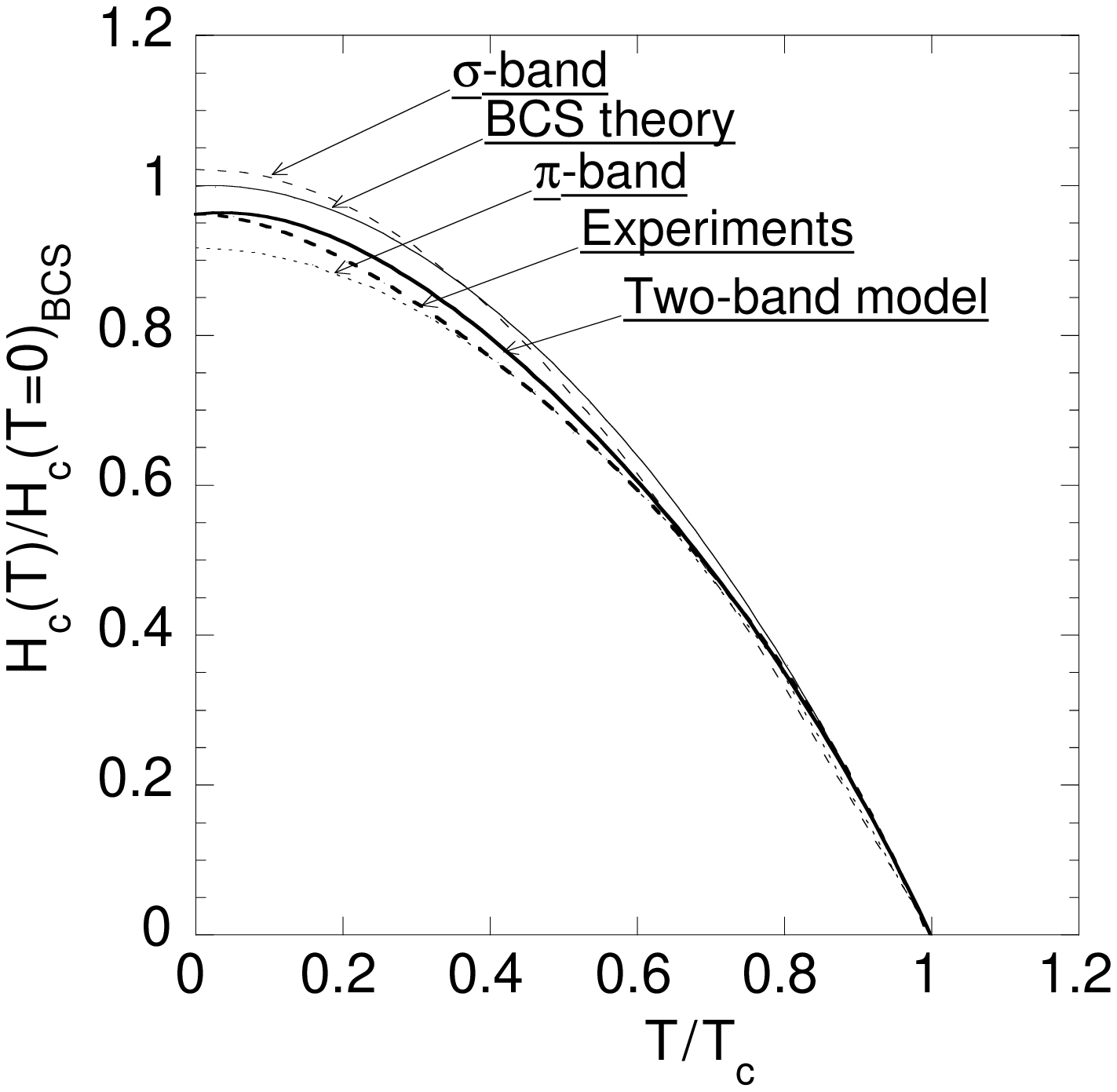}
\caption{
Thermodynamic critical magnetic field $H_c(T)$ normalized by $H_c(T=0)_{BCS}$.
The bold dashed curve indicates data from ref.\cite{bou01} and the bold solid
curve indicates those obtained using the
present two-band anisotropic model.  The thin solid curve indicates the BCS results.
The results for the $\sigma$-band (ab-plane anisotropic) and the $\pi$-band
(prolate form) are also shown.
}
\label{hc}
\end{figure}

The R-T method provides us reliable data of spectroscopy in far-infrared
region for which comparison between the experimental data and theoretical
analysis is possible.  In the R-T method both the reflectance spectra $R(\omega)$ 
and the transmittance spectra $T(\omega)$ are measured experimentally from
which a set of coupled equations are followed describing the transmittance and
reflectance of a thin film on a substrate.  The coupled equations are solved
numerically by the Newton method to determine the optical conductivity.
This method is free from the difficulties in the infrared region which occur
commonly in the conventional method employing a Kramers-Kronig transformation. 
In Fig.\ref{sigma1Ex}, we show the real part of the optical conductivity obtained from
the R-T method at $T=3.4$K and $T=30$K.
In Fig.\ref{sigma2} we show the observed data and theoretical curves at $T=0$ for
$2\Delta=60$ and 70 cm$^{-1}$.  The experimental data 
$\sigma(3.4K)/\sigma(30K)$ normalized by the normal state values at $T=30$K 
are shown in Fig.\ref{sigma2}.  It is obvious from the experimental results that there 
is no evidence of a true gap, which is suggestive of anisotropic 
superconducting gap, since the spectral weight of conductivity should vanish for 
$\omega\le 2\Delta$ at $T=0$ in conventional isotropic superconductors.
It is also shown in Fig.\ref{sigma2} that they are well fitted by the curve with 
$2\Delta=60$ cm$^{-1}$, which is consistent with the value estimated by
scanning tunneling spectroscopy measurements.\cite{kas98}

Transmission curve is also presented in Fig.\ref{tstn}, where $T_S/T_N$, the ratio of
the transmission in the superconducting to that in the normal state, is the
experimentally measured quantity.  The following expression for $T_S/T_N$
is employed to determine the transmission curve theoretically,\cite{glo57}
\begin{equation}
\frac{T_S}{T_N}= \frac{1}{[T_N^{1/2}+(1-T_N^{1/2})(\sigma_1/\sigma_n)]^2+
[(1-T_N^{1/2})(\sigma_2/\sigma_n)]^2},
\end{equation}
where $\sigma_1$ and $\sigma_2$ are real and imaginary parts of the
conductivity $-(c/\omega) K({\bf q},\omega)$ for ${\bf q}\rightarrow 0$, respectively.
Here we use the formula obtained from the two-fluid model for $\sigma_2$.
$T_N$ is determined as $T_N\simeq 0.05$ from the
expression for the ratio of the power transmitted with a film to that with
no film given as
\begin{equation}
T_N= 1/[1+\sigma_n d\frac{Z_0}{n+1}]^2 .
\end{equation}
Here $d$ is the film thickness, $n$ is the index of refraction of the substrate,
and $Z_0$ is the impedance of free space.  We have assigned the following
values;  $d=4\times 10^{-6}$cm, $n=3.13$, $Z_0=377 \Omega$ and 
$\sigma_n\approx 10^4 \Omega^{-1}$cm$^{-1}$ and the Drude width is approximately
equal to $\Delta$
$\sigma_n$ is approximately given by the value at $\omega=0$.
Obviously the $\omega$-dependence of measured transmittance agrees with the 
theoretical curve for
$2\Delta=60$cm$^{-1}$ as shown in Fig.\ref{tstn}.
An agreement between the observed quantities and theoretical curve is
remarkable, which should be compared to the isotropic BCS prediction
calculated from the Mattis-Bardeen equations.\cite{cho96}

\section{Two-Band Anisotropic Superconductivity in Magnesium Diboride}
After the discovery of 39 K superconductivity in MgB$_2$\cite{nag01}, much attention 
has been focused on the study of its nature.
An $s$-wave superconductivity (SC) was established by experiments such as
coherence peak in $^{11}$B nuclear relaxation rate\cite{kot01} and its exponential
dependence at low temperatures\cite{yang01,man02}.
An isotope effect has suggested phonon-mediated $s$-wave superconductivity\cite{bud01}.In contrast to its standard properties, there have been several reports indicating
unusual properties of the superconductivity of MgB$_2$.
Two different superconducting gaps have been reported:
a gap much smaller than the expected BCS value and that comparable
to the BCS value given by $2\Delta=3.53k_BT_c$.
Their ratio is estimated to be $\Delta_{min}/\Delta_{max}\sim 0.3-0.4$ using several
experiments\cite{yang01,tsu01,sza01,che01,giu01,bou01}.
It is also reported that the specific-heat jump and the critical magnetic field
are reduced compared to the $s$-wave BCS theory\cite{yang01,bou01}.
A strongly anisotropic upper critical field in
$c$-axis-oriented MgB$_2$ films and single crystals of
MgB$_2$ is also observed\cite{lim01,xu01,ang02}.

\begin{table}
\caption{Anisotropic parameters in the SC gap function used to fit several physical
quantities.  The upper four rows are for the single-SC gap model and the last row is
for
the two-band anisotropic model for comparison.
The cross indicates that we cannot fit experimental data by the corresponding
$z$ factor.  $\Delta$ in the column $H_{c2}$ indicates that
experiments are explained qualitatively but not quantitatively.\cite{haa01}
The big circle show that we can fit the data using the same parameters
in the column $\sigma_1$.
The effect of $\sigma$-band anisotropy is small.
}
\label{table1}
\begin{center}
\begin{tabular}{cccccccc}
\hline
 & $z$ & $\sigma_1$ & $T_S/T_N$ & $\Delta C$ & $H_c(0)$ &
$H_{c2}$  \\
\hline
Cigar-type & $1+a{\rm cos}(2\theta)$ & $a\sim 0.5$ & $\sim 0.3$ & $\sim 0.3$
& $\sim 0.07$      & {\large $\times$} \\
Pancake    & $1-a'{\rm cos}(2\theta)$& $a'\sim 0.6$& {\large $\times$} & $\sim 0.5$
& {\large $\times$} & $\Delta$  \\
Pancake    & $1-b{\rm cos}^2(\theta)$& $b\sim 0.75$& {\large $\times$} & $\sim 0.66$
& $\sim 0.08$      & $\Delta$  \\
In-plane & $1+c{\rm cos}(6\phi)$     & $c\sim 0.5$ & $\sim 0.3$ & $\sim 0.3$
& {\large $\times$} &   \\
\hline
Two-band & ($\sigma$ band) & $c <  0.3$ & $\bigcirc$ & $\bigcirc$ & $\bigcirc$ &  \\
         & ($\pi$ band)    & $a\sim 0.3$ &  & &  & \\
\hline
\end{tabular}
\end{center}
\end{table}

The unusual properties of MgB$_2$ suggest an anisotropic $s$-wave superconductivity
or a two-band superconductivity.
The band structure calculations predicted multibands originating from
$\sigma (2p_{x,y})$ and $\pi (2p_z)$ bands.\cite{kor01}
In the ARPES measurements performed in single crystals of MgB$_2$ three distinct
dispersions approaching the Fermi energy were reported.\cite{uch02}

There have been several studies on the anisotropy of a superconducting
gap\cite{haa01,mir02,nak02,tew02,bou02}.
The two-gap model is shown to consistently describe the specific heat\cite{nak02,bou02}and the upper critical field $H_{c2}$\cite{mir02} with the adoption of the effective
mass approach.

Here we show that this material is described by two order parameters attached to
$\sigma$- and $\pi$-bands.  Two order parameters further must have different 
anisotropy to explain the experimental results consistently.
In this paper, we examine optical properties and thermodynamics to determine the
{\bf k}-dependence of the gaps.
We show that the optical transmittance, conductivity, specific-heat jump,
and thermodynamic critical field $H_c$ are well described by a two-band
superconductor model with different anisotropies in ${\bf k}$-space.
The symmetry in ${\bf k}$-space is determined in order to explain these
experiments consistently.

The optical conductivity for anisotropic s-wave SC is
investigated and compared with available data for MgB$_2$.
A simple angle-dependent generalization of the Mattis-Bardeen formula\cite{mat58}
is used to calculate the optical conductivity.
The density of states $N(\epsilon)=\epsilon/\sqrt{\epsilon^2-\Delta^2}$ is
generalized to
$N(\epsilon)=\langle{\rm Re}\epsilon/\sqrt{\epsilon^2-\Delta_{{\bf k}}^2}\rangle_k$,
where the bracket indicates the average over the Fermi surface.
We employed the following formula at T=0:
\begin{equation}
\frac{\sigma_{1s}}{\sigma_{1n}}(\omega)
=\frac{1}{\omega} \int_0^{\omega}dE [ N(E)N(\omega-E)
- \langle {\rm Re} \frac{\Delta_{\bf k}}{\sqrt{E^2-\Delta_{\bf k}^2}} \rangle _k
\langle {\rm Re} \frac{\Delta_{\bf k'}}{\sqrt{(\omega-E)^2-\Delta_{{\bf k'}}^2}}
\rangle _{k'} ],
\label{sigma1s}
\end{equation}
\begin{eqnarray}
\frac{\sigma_{2s}}{\sigma_{1n}}(\omega)&=& -\frac{1}{\omega}
\int^{\omega+\Delta_{max}}_{\Delta_{min}}dE[ \langle{\rm Re}
\frac{E}{\sqrt{E^2-\Delta_k^2}}\rangle_k \langle{\rm Re}
\frac{\omega-E}{\sqrt{\Delta_{k'}^2-(\omega-E)^2}}\rangle_{k'}\nonumber\\
&-& \langle{\rm Re}\frac{\Delta_k}{\sqrt{E^2-\Delta_k^2}}\rangle_k \langle{\rm Re}
\frac{\Delta_{k'}}{\sqrt{\Delta_{k'}^2-(\omega-E)^2}}\rangle_{k'}]. 
\end{eqnarray}
We here mention that if the samples are clean and belong to the category of London
superconductors, we must use the formulas in the London limit.
For a clean superconductor, it seems better to use $\sigma_{2s}$ for the
two-fluid model\cite{shi03}.
The optical data that we will consider here exhibit behaviors explained by
the conventional formulas of Mattis and Bardeen.
The anisotropic order parameters considered in this paper are:
\begin{eqnarray}
\Delta_{c1}({\bf k})&=& \Delta (1+a{\rm cos}(2\theta)),\nonumber\\
\Delta_{c2}({\bf k})&=& \Delta (1-b{\rm cos}^2(\theta)),\nonumber\\
\Delta_{ab}({\bf k})&=& \Delta (1+c{\rm cos}(6\phi)).
\end{eqnarray}
Here, $\theta$ and $\phi$ are the angles in the polar coordinate where $\theta$ is
the polar angle with respect to the c-axis.  The parameters $a$, $b$ and $c$
determine the anisotropy.
$\Delta_{c1}$ is a prolate form gap for $a>0$ and
is oblate for $a<0$.
$\Delta_{c2}$ ($b>0$) shows the same anisotropy as $\Delta_{c1}$ for $a<0$.
$\Delta_{ab}({\bf k})$ indicates an anisotropy in the $ab$-plane; the SC gap may
possibly be anisotropic in the plane since the 2D-like Fermi surface has a
hexagonal symmetry.\cite{kor01}
The integral in eq.(\ref{sigma1s}) is evaluated numerically by writing the average 
over the
Fermi surface with elliptic functions.  For example, for 
$\Delta_{c1}({\bf k})= \Delta (1+a{\rm cos}(2\theta))$, the average over the Fermi
surface for $0<a<1$ is given by
\begin{eqnarray}
\langle {\rm Re}\frac{\omega}{\sqrt{\omega^2-\Delta_{c1}({\bf k})^2}}\rangle&=&
\frac{1}{2}\sqrt{\frac{\omega}{a\Delta}}F\left(\frac{\pi}{2},k\right),~~(1-a)\Delta
\le\omega\le (1+a)\Delta,\nonumber\\
&=& \frac{1}{2}\sqrt{\frac{\omega}{a\Delta}}F\left(\gamma,k\right),~~
(1+a)\Delta<\omega,
\end{eqnarray}
where $k^2=(\omega-(1-a)\Delta)/(2\omega)$,
$\gamma={\rm sin}^{-1}\sqrt{4a\Delta\omega/[(\omega-(1-a)\Delta)(\omega+(1+a)\Delta)]}$
and $F(\gamma,k)$ is the elliptic integral of the first kind.

\begin{table}
\caption{Several physical quantities obtained by the two-band model
with $c\sim 0.33$ ($\sigma$ band) and $a\sim 0.33$ ($\pi$ band).
}
\label{table2}
\begin{center}
\begin{tabular}{ccccc}
\hline
 & $w_{\sigma}/w_{\pi}$ & $\Delta_{min}/\Delta_{max}$ &
$\frac{\Delta C(T_c)}{\Delta C(T_c)_{BCS}}$  & $\frac{H_c(0)}{H_c(0)_{BCS}}$   \\
\colrule
Two-band & 0.45/0.55 & $\sim 0.35$ & $\sim 0.82$ & $\sim 0.95$  \\
Exp.     & 0.45/0.55  & $0.3-0.4$ & $0.76-0.92$
& 0.96 \\
\hline
\end{tabular}
\end{center}
\end{table}

First, we examine a one-band anisotropic model and show that the one-band model
is insufficient to understand consistently
optical and thermodynamic behaviors.
In Fig. \ref{tstn1} the transmission $T_{S}$ at
$T=0$ is shown as a function of the frequency $\omega$.\cite{yan03b}
We again employ the following phenomenological expression for
$T_S/T_N$\cite{yana01,glo57},
\begin{equation}
\frac{T_S}{T_N}= \frac{1}{[T_N^{1/2}+(1-T_N^{1/2})(\sigma_1/\sigma_n)]^2+
[(1-T_N^{1/2})(\sigma_2/\sigma_n)]^2},
\end{equation}
where $\sigma_1$ and $\sigma_2$ are real and imaginary parts of the optical
conductivity, respectively.
$T_N$ is determined from the
expression for the ratio of the power transmitted with a film to that transmitted
without a film given as
$T_N= 1/[1+\sigma_n d Z_0/(n+1)]^2$.
Here, $d$ is the film thickness, $n$ is the index of refraction of the
substrate,
and $Z_0$ is the impedance of free space.
We have assigned the following
values:  $d= 10^{-6}$ cm, $n\approx 3$, $Z_0=377 \Omega$, and
$\sigma_n\approx 8\times 10^3 \Omega^{-1}$cm$^{-1}$.
Then we obtain $T_N\simeq 0.014$.
The theoretical curves for $T_S/T_N$ are shown in Fig. \ref{tstn1}; they have peaks
near $\omega\sim 2\Delta_0$.
For the oblate, its peak shows an increase only twice the normal state
value, while the prolate and ab-plane anisotropies show more than
twofold increases.   The experiments show an approximately 2.5-fold
increase\cite{kai02}
which supports the prolate or ab-plane anisotropic symmetry.
However, the
temperature dependence of the ratio $H_{c2}^{ab}/H_{c2}^c$, which increases as the
temperature decreases\cite{ang02}, indicates that $\Delta({\bf k})$
has an oblate form instead of a prolate form\cite{haa01} in contrast to $T_S/T_N$.
It is also difficult to describe the thermodynamic quantities such as the
specific-heat jump at $T=T_c$ and the thermodynamic critical magnetic field $H_c$
within the single-gap model consistently.
The specific-heat jump at $T_c$ is given by
\begin{equation}
\frac{\Delta C(T_c)}{\gamma_C T_c}=\frac{12}{7\zeta (3)}
\frac{\langle z^2\rangle ^2}{\langle z^4\rangle },
\end{equation}
where $\gamma_C$ is the specific-heat coefficient and $z$ is an anisotropic factor
of the gap function.  $\langle z^n\rangle$ is the average of $z^n$ over the
Fermi surface.
In Fig. \ref{spec} the specific-heat-jump ratio vs anisotropy ($a$ or $c$) is shown.
The experiments indicate that this value is in the range of 0.76$\sim$
0.92\cite{yan01,bou01}; the fitting parameters must be $a\sim 0.3$,
$-a\sim 0.5$ and $c\sim 0.3$ for the prolate, oblate and ab-anisotropic types,
respectively.
We must assign different values to parameters $a$ and $c$ in order to explain
the thermodynamic critical magnetic field $H_c$.
The ratio of $H_c(T=0)$ to the BCS value is given as
\begin{equation}
\frac{H_c(T=0)^2}{\gamma_CT_c^2}=\frac{6\pi}{{\rm e}^{2\gamma}}\langle z^2\rangle=
5.94\langle z^2\rangle.
\end{equation}
Thus to be consistent with the experimental results\cite{bou01},
$\langle z^2\rangle$ should be less than 1;
$a$ should be small, $a\sim 0.07$, for the prolate form, and the ab-plane anisotropic
and oblate forms $(a<0$) are ruled out since $\langle z^2\rangle >1$.
In Table I, we summarize the status for the single-gap anisotropic $s$-wave model
applied to MgB$_2$. As shown here, it is difficult to understand the physical 
behaviors measured using several experimental methods consistently within the 
single-gap model.

Here, a two-band model with two different anisotropies is investigated.
We assume that the hybridization between $\sigma$ and $\pi$ bands is negligible,
and that
the optical conductivity is given by
\begin{equation}
\sigma= w_{\sigma}\sigma_{\sigma}+w_{\pi}\sigma_{\pi},
\end{equation}
where $\sigma_{\sigma}$ and $\sigma_{\pi}$ denote the contributions from $\sigma$-
and $\pi$-bands, respectively.
For the case of isotropic two gaps, $\sigma_1$ must have a shoulder-like structure
which appear as an addition of two contributions from the two bands, if the
magnitudes of two SC gaps are different.
The experimental data of $\sigma_1$, however, does not have such a sharp structure
(see Fig. \ref{s1}).\cite{kai02,lee02}  Therefore we must take account of 
anisotropies for
the two-band model.  We assume the in-plane anisotropy for the two-dimensional-like
$\sigma$-band, while we assign the three-dimensional anisotropy to $\pi$-band
where the prolate and oblate forms are examined.

The transmission $T_S/T_N$ in Fig. \ref{tstn2} shows that the theoretical
curve is in good agreement with the experimental curve.
The optical conductivity is also described well by the two-band model as shown
in Fig. \ref{s1}.
We assign the following parameters to the best fit model in Figs. \ref{tstn2} and 
\ref{s1};
the $\sigma$-band has ab-plane anisotropy with $c\approx$ or less than 0.33 and
the $\pi$-band has the prolate form gap (cigar type) with $a\approx 0.33$.
The ratio of the weight of the $\sigma$-band to that of the $\pi$-band is 0.45/0.55,
which
agrees with penetration depth\cite{man02} and band structure calculations\cite{bel01}.
The ratio of the minimum gap to the maximum gap is 0.35, which is in the range of
previously reported experimental values.\cite{tsu01,giu01}
Let us mention that the effect of $\sigma$-band anisotropy is small for the
transmission $T_S/T_N$.

In Fig. \ref{hc} the thermodynamic critical magnetic field $H_c(T)$ is shown for the
single-band and two-band models with available data.\cite{bou01}
We have simply assumed that the total free energy is given by the sum of two
contributions from $\sigma$- and $\pi$-bands:
$\Omega=w_{\sigma}\Omega_{\sigma}+w_{\pi}\Omega_{\pi}$.
The experimental behavior is well explained by the two-band anisotropic model
using the same parameters as those for $T_S/T_N$ and $\sigma_{1s}/\sigma_{1n}$.
We show several characteristic values obtained from the two-band model
in Table II.  

Let us mention here that the two-gap model shows consistency concerning
other physical quantities.
Results of
analyses of $H_{c2}$ and specific heat using the effective mass approach are
consistent with those obtained using the two-band model.\cite{mir02,nak02,bou02}
It has been reported that the increasing nature of $H_{c2,ab}/H_{c2,c}$ with
decreasing temperature is explained by the two-Fermi surface model.\cite{mir02}
The specific-heat coefficient $\gamma$ in magnetic fields seems consistent
with that of the multiband superconductor.\cite{nak02,bou02}

\section{Summary}
We have discussed the optical properties in unconventional superconductors.
Theoretical aspects of the conductivity were discussed in detail from the linear
response theory to the formula in the London limit.
We have presented a new method (R-T method) to measure $\sigma(\omega)$ in the 
far-infrared
region from reflectance and transmittance data without the use of the Kramers-Kronig 
transformations.  This method provides a method to obtain the far-infrared
properties more precisely compared to the conventional method.
The conductivity sum rule is discussed briefly.  It has been reported that the sum
rule is satisfied for the optical conductivity spectra of NbN$_{1-x}$C$_x$ that is
a typical conventional superconductor.

We have successfully made a comparison between experiments and theory
for the optical conductivity of Nd$_{2-x}$Ce$_x$CuO$_4$ in the far-infrared 
region.
We have shown that there is a reasonable agreement between the optical
conductivity $\sigma_1(\omega)$ observed by the R-T method and theoretical 
analysis without
adjustable parameters except the superconducting gap.  An estimate of
60$\sim$70 cm$^{-1}$ for the superconducting gap is consistent from both
the experimental and theoretical aspects.
The far-infrared optical conductivity suggests that the superconducting gap
of electron-doped Nd$_{2-x}$Ce$_x$CuO$_4$ is unconventional one with nodes
on the Fermi surface.
The anisotropic nature of electron-doped superconductors is consistent
with the recent research performed for the one-band and three-band Hubbard
models.\cite{yam98,kur99,yan01,yan02,yan03}
If the superconducting gap is anisotropic for the electron-doped
superconductors, there is a possibility that both the hole-doped and 
electron-doped cuprates superconductors are governed by a same superconductivity
mechanism.

We have also examined the transmittance,  optical conductivity,
specific-heat jump and thermodynamic critical magnetic field $H_c$ of MgB$_2$ based
on the two-band anisotropic $s$-wave model.
This material is described by two order parameters attached to $\sigma$- and
$\pi$-bands, respectively, which, moreover, have further anisotropy.
We have shown that the two-gap model with different anisotropy in ${\bf k}$-space can
explain the experimental results consistently. 

\section{acknowledgment}
We express our sincere thanks to our coworkers: E. Kawate, S. Kimura, S. Kashiwaya, 
A. Sawa and S. Kohjiro.  We thank Professor K. Maki for comments on the London
superconductor.

\end{document}